\documentclass[final,5p,times,twocolumn,authoryear]{elsarticle}

\usepackage{amssymb}
\usepackage{lipsum}
\usepackage{amsmath}
\usepackage[colorlinks=true,citecolor=blue]{hyperref}
\usepackage{graphicx,graphics,color}
\usepackage{ulem}
\usepackage{color}

\usepackage{lineno}

\newcommand\chandra{{\it Chandra~}}

\newcommand\Hunit{\ifmmode {\rm~km\ s}^{-1}\ {\rm Mpc}^{-1}
        \else ~km s$^{-1}$ Mpc$^{-1}$\fi}
\newcommand\ctssec{\ifmmode {\rm~count\ s}^{-1} \else ~count s$^{-1}$\fi}
\newcommand\ergsec{\ifmmode {\rm~erg\ s}^{-1} \else
        ~erg s$^{-1}$\fi}
\newcommand\ergs{\ifmmode {\rm~erg\ s}^{-1} \else
        ~erg s$^{-1}$\fi}
\newcommand\funit{\ifmmode {\rm~erg\ s}^{-1}\;{\rm cm}^{-2} \else
        ~ergs s$^{-1}$ cm$^{-2}$\fi}
\newcommand\phflux{\ifmmode {\rm~photon\ s}^{-1}\;{\rm cm}^{-2}
        \else   ~photon s$^{-1}$ cm$^{-2}$\fi}
\newcommand\efluxA{\ifmmode {\rm~erg\ s}^{-1}\;{\rm cm}^{-2}\;{\rm
        \AA}^{-1} \else ~erg s$^{-1}$ cm$^{-2}$ \AA$^{-1}$\fi}
\newcommand\efluxHz{\ifmmode {\rm~erg\ s}^{-1}\;{\rm cm}^{-2}\;{\rm
        Hz}^{-1} \else ~erg s$^{-1}$ cm$^{-2}$ Hz$^{-1}$\fi}
\newcommand\cc{\ifmmode {\rm~cm}^{-3} \else cm$^{-3}$\fi}
\newcommand\FWHM{\ifmmode {\rm~FWHM} \else ${\rm~FWHM}$\fi}
\newcommand\Msun{\ifmmode M_{\odot} \else $M_{\odot}$\fi}
\newcommand\Zsun{\ifmmode Z_{\odot} \else $M_{\odot}$\fi}
\newcommand\Lsun{\ifmmode L_{\odot} \else $L_{\odot}$\fi}

\newcommand\hbeta{\ifmmode {\rm H}\beta \else H$\beta$\fi}
\newcommand\Kalpha{\ifmmode {\rm K}\alpha \else K$\alpha$\fi}
\newcommand\nh{\ifmmode N_{\rm H} \else N$_{\rm H}$\fi}

\newcommand{\mnras}{MNRAS}

\def\H2{\hbox{H$_{2}$}}

\newcommand{\arcsec}{\ensuremath{^{\prime\prime}}}

\newcommand{\degr}{\ensuremath{\hbox{$^\circ$ }}}

\journal{New Astronomy}

\begin{document}

\begin{frontmatter}



\title{Sloshing Cold Fronts in Galaxy Cluster Abell 2566}


\author[first]{S. K. Kadam}
\affiliation[first]{organization={School of Physical Sciences, Swami Ramanand Teerth Marathwada University},%Department and Organization
 %          addressline={}, 
            city={Nanded},
            postcode={431606}, 
            state={Maharashtra State},
            country={India}}

\author[second]{S. S. Sonkamble}
\affiliation[second]{organization={Centre for Space Research, North-West University},%Department and Organization
%           addressline={}, 
            city={Potchefstroom},
            postcode={2520}, 
            state={North West Province},
            country={South Africa}}

\author[third]{N. D. Vagshette}
\affiliation[third]{organization={Department of Physics and Electronics, Maharashtra Udayagiri Mahavidyalay},%Department and Organization
%           addressline={}, 
            city={Udgir},
            postcode={413517}, 
            state={Maharashtra State},
            country={India}}

\author[first]{M. K. Patil}\ead{patil@associates.iucaa.in}

\begin{abstract}
This paper presents properties of the intracluster medium (ICM) in the environment of a cool core cluster Abell~2566 (redshift $z$ = 0.08247) based on the analysis of 20~ks \chandra X-ray data. 2D imaging analysis of the \textit{Chandra} data from this cluster revealed spiral structures in the morphology of X-ray emission from  within the central 109 kpc formed due to gas sloshing. This analysis also witness sharp edges in the surface brightness distribution along the south-east and north-west of the X-ray peaks at 41.6 kpc and 77.4 kpc, respectively. Spectral analysis of 0.5 - 7 keV X-ray photons along these discontinuities exhibited sharp temperature jumps  from 2.3 to 3.1 keV and 1.8 to 2.8 keV, respectively, with consistency in the pressure profiles, implying their association with cold fronts due to gas sloshing of the gas. Further confirmation for such an association was provided by the deprojected broken power-law density function fit to the surface brightness distribution along these wedge shaped sectorial regions. This study also witness an offset of 4.6\arcsec\, (6.8 kpc) between the BCG and the X-ray peak, and interaction of the BCG with a sub-system in the central region, pointing towards the origin of the spiral structure due to a minor merger. 
\end{abstract}



\begin{keyword}
%% keywords here, in the form: keyword \sep keyword, up to a maximum of 6 keywords
galaxies: clusters: general; galaxies: clusters: individual Abell~2566; X-rays: galaxies: clusters; galaxies: clusters: intracluster medium
\end{keyword}



\end{frontmatter}

%\tableofcontents

%\linenumbers

\section{Introduction}
\label{introduction}

Being the largest gravitationally bound systems in the universe, detailed study of physical properties of the intracluster medium (ICM) in the environment of galaxy clusters provide systematic information on the evolution of its host \citep{1980ApJ...236..351D,2000MNRAS.318..889W}. It is likely that the impact of the thermodynamic disturbance due to merger like episodes on the thermal state of the ICM is imprinted in its complex morphology in the form of substructures like, cold fronts \citep{2000ApJ...541..542M,2001ApJ...551..160V}, shocks \citep{2022MNRAS.509.5821R} and spiral structures \citep{2010ApJ...710.1776J}. Among these, cold fronts and shocks have been witnessed as arcs around the cool cores in several of the relaxed galaxy clusters \citep{2003ApJ...590..225C,2005ApJ...618..227T,2006ApJ...650..102A,2015Ap&SS.359...61S,2017MNRAS.466.2054V}. The bulk motions and shocks induced by merger like events may generate turbulence that heat up the surrounding ICM \citep{2015SSRv..188..141B}. Particularly, an off-axis interaction of a perturber (galaxy) while it passes near the central system may induce an offset in the core of cluster as it passes around the centre of the potential well and may result in the formation of spiral-like structures in the surface brightness distribution \citep{2001ApJ...562L.153M,2006ApJ...650..102A}.  

High resolution X-ray imaging studies using data from \chandra telescope have revealed that the most common features of merging events seen in galaxy groups and clusters is `sloshing'. The sloshing of gas in such merger events have yielded the spiral-like contact discontinuities, commonly known as cold fronts,  in the surface brightness distribution \citep[][]{2003ApJ...583L..13D,2010ApJ...710.1776J,2016MNRAS.457...82S,2019ApJ...875...65C,2011ApJ...743...15M,2007PhR...443....1M}.   Such spiral structures have also been witnessed in several other galaxy clusters e.g., Abell 2052 \citep{2011ApJ...737...99B}, RXJ2014.8-2430 \citep{2014MNRAS.441L..31W}, Centaurus \citep{2016MNRAS.457...82S}, Abell 1763 \citep{2018ApJ...868..121D}, Abell 795 (Kadam et al. 2023 under review), etc.

Other observable characteristics of the cold fronts are provided by the spectral analysis of the X-ray emission from galaxy groups and cluster environments. At the locations of discontinuities due to the cold fronts, higher density side of the discontinuity is relatively cooler than that on the fainter side \citep[e.g.,][]{2000ApJ...541..542M,2001ApJ...551..160V}. Though a jump in the temperature and density of gas is evident across such a discontinuity, the pressure profile remains typically constant e.g., PKS 0745-191 \citep{2014MNRAS.444.1497S}, Abell 2390 \citep{2015Ap&SS.359...61S}, the Perseus cluster \citep{2019MNRAS.483.1744I}.  Further studies of merging clusters using high resolution data from \chandra observatory have revealed that the sloshing of gas yields multiple cold fronts at different locations, growing over a time period and combining into a spiral shape \citep{2011MNRAS.413.2057R}. Although cold fronts have been evidenced in several of the galaxy groups and clusters, their association with the sloshing is apparent only in handful systems \citep[e.g.,][]{2001ApJ...555..205M,2007ApJ...671..181D,2009AIPC.1201..237G,2009ApJ...700.1404R}. Therefore, a detailed study of cold fronts, particularly at the locations of sloshing effects, is called for as it provides a  potential probe for understanding the turbulence, gas heating, ICM enrichment, etc. \citep{2009ApJ...696.1431R}.

%%%%%%%%%%%%%%%%%%%%%%%%%%%%%%%%
\begin{figure*}
\centering
\includegraphics[scale=0.44]{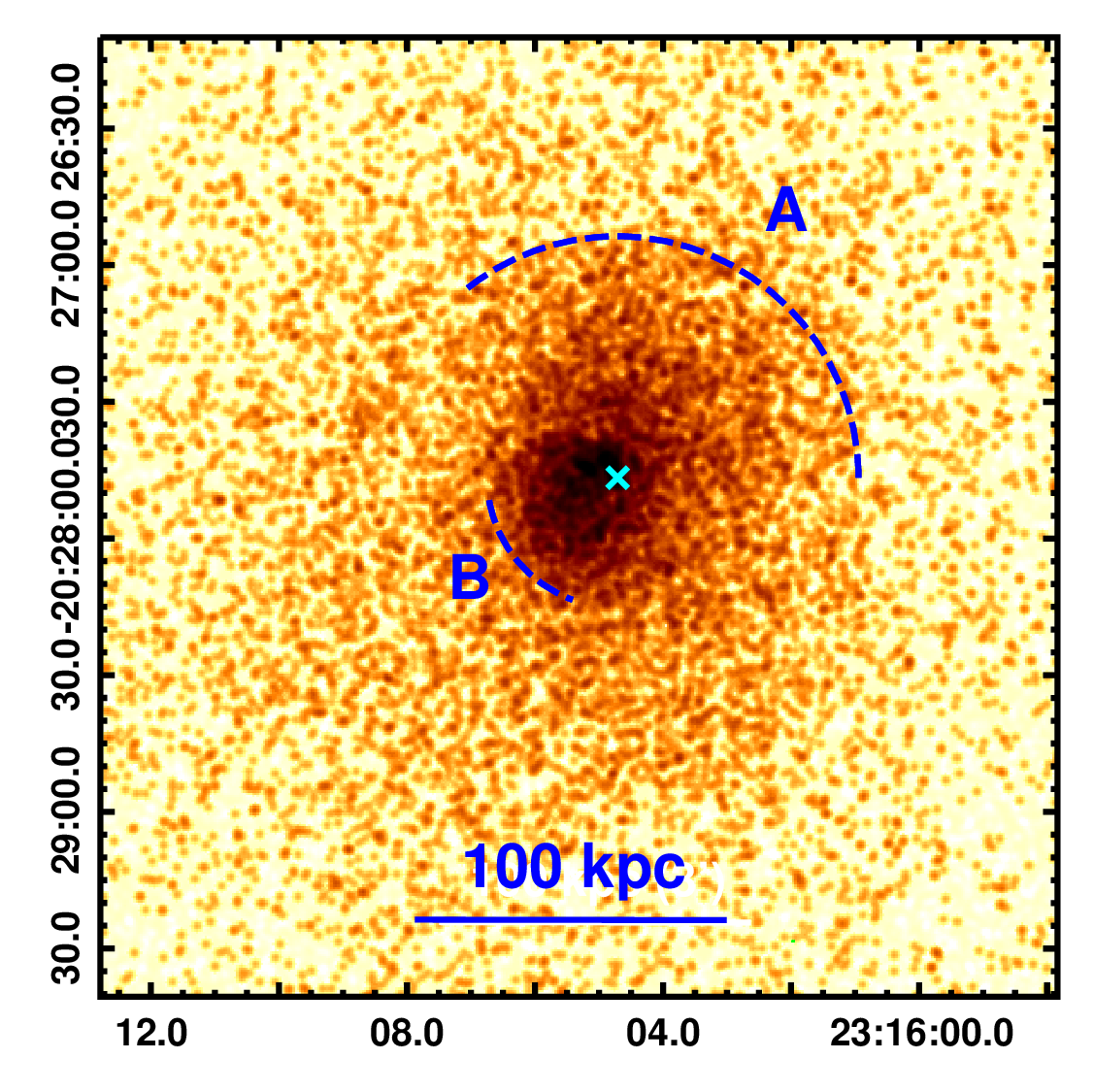}
\includegraphics[scale=0.44]{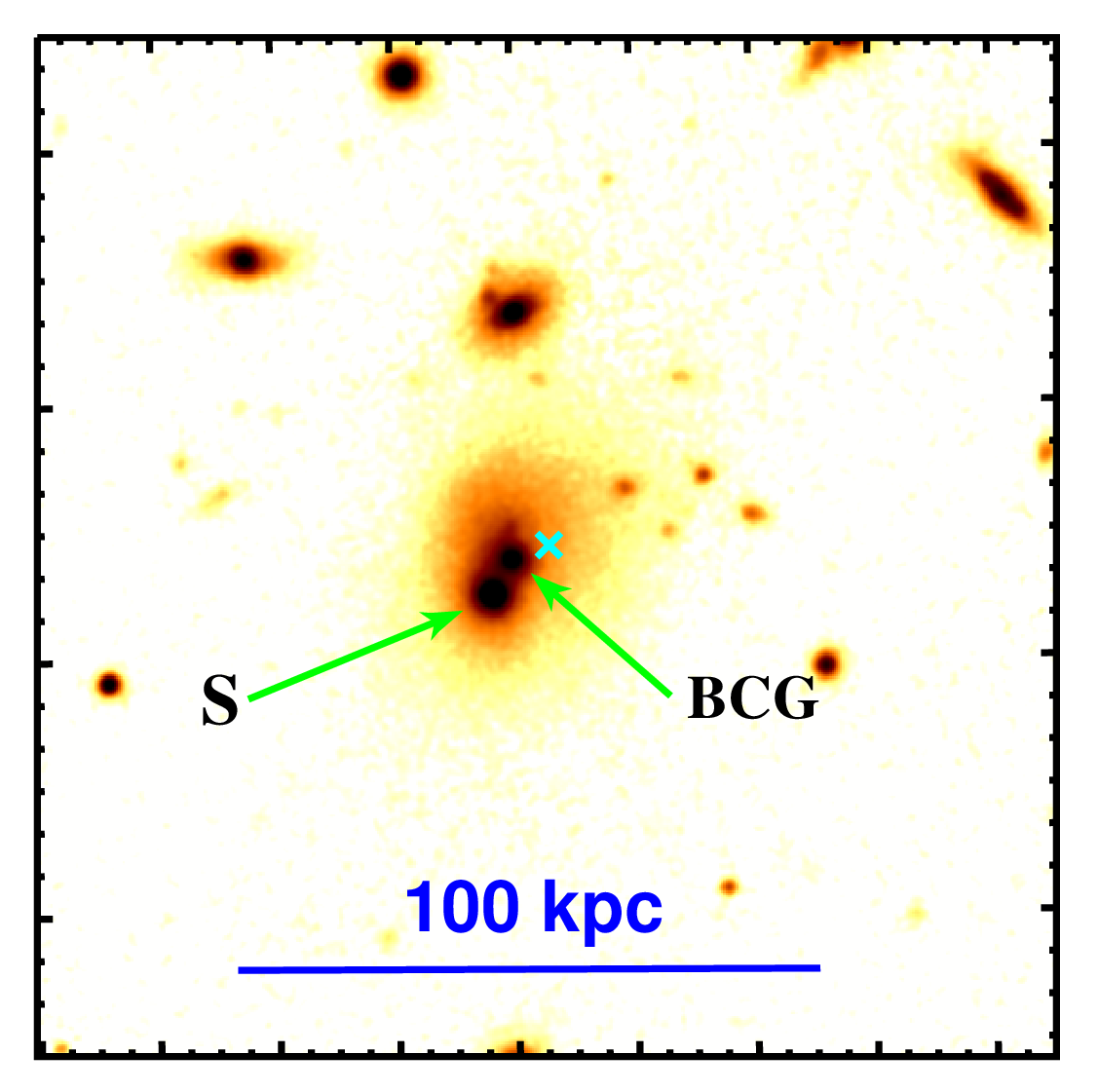}
\caption{{\it Left panel}: Background-subtracted, exposure-corrected, point source removed 0.5 - 3 keV 3.5 $\times$ 3.5 arcmin \textit{Chandra} X-ray image of A2566. For better visibility this image has been smoothed with 3$\sigma$ Gaussian kernel. X-ray peak of the cluster emission is marked by cyan cross. The sectorial arc shaped regions used for deriving surface brightness profile (see Section~\ref{sb}) and spectral analysis (see Section~\ref{spectra_ana})  are marked as A and B. {\it Right panel}: Pan-STARRS 1 $\times$ 1 arcmin r-band image indicating position of the BCG A2566 along with its nearby member S.}
\label{raw}
\end{figure*}
%%%%%%%%%%%%%%%%%%%%%%%%%%%%%%%%%%%%%%%%%%%

This paper present evidences of gas sloshing in the core of Abell 2566 (hereafter A2566) along with a pair of cold fronts in its environment employing \textit{Chandra} X-ray observations and radio data. A2566 is a massive cluster with $M_{500}=2.17\pm0.09\times10^{14}$ \Msun\, \citep{2019MNRAS.483..540L}. The line of sight velocity measured using the cluster emission centered on the H$\alpha$+[NII] is reported to be 325$\pm$22 km s$^{-1}$ \citep{2016MNRAS.460.1758H}. This system also hosts a radio source with a total flux density of 23 mJy at 1400 MHz \citep{1997ApJS..108...41O}.  The paper is organized as follows. Section~\ref{sec2} provides a brief explanation on the data preparation,  while Section~\ref{sec3} presents results derived from the X-ray imaging and spectroscopic study. Section~\ref{disc} provides discussion on the results and its association with other components and Section~\ref{con} summaries our findings.
This paper has taken the cosmological parameters from \cite{2007ApJS..170..377S}, which are H$_0$= 73 km s$^{-1}$ Mpc$^{- 1}$, $\Omega_M$ = 0.27, and $\Omega_{\Lambda}$ = 0.73. This means 1 arcsec is equivalent to 1.48 kpc at a redshift of 0.08247. The errors quoted in the spectral analysis are at 68$\%$ confidence level and the reported metallicities were measured relative to those in the table of \citet{1998SSRv...85..161G}.

%%%%%%%%%%%%%%%%%%%%%%%%%%%%%%%%%%%%%%%%%%%%%%%%%%%%%%%%%%%%%%%%%%%%%%%%%%%%%%
\begin{table*}
%\scriptsize
\centering
\caption{\textit{Chandra} X-ray observations of A2566}
\begin{tabular}{ccccccccr}
\hline
\textit{ObsID} & Instrument & Date Obs & Data Mode & PI & Exposure (ks) & Cleaned Exposure (ks) \\
\hline
19595 & ACIS-S & 30 Sept. 2016 & VFAINT & A. Edge & 20 & 19.4 \\
\hline
\end{tabular}
\label{obs}
\end{table*}
%%%%%%%%%%%%%%%%%%%%%%%%%%%%%%%%%%%%%%%%%%%%%%%%%%%%%%%%%%%%%%%%%%%%%%%%%%%%%%

\section[2]{Observations and Data Preparation}
\label{sec2}
Level 1 event files of A2566 acquired using \chandra ACIS-S in \textit{VFAINT} mode (ObsID: 19595; Table~\ref{obs}) were reprocessed using \textit{CIAO 4.14} and calibration files \textit{CALDB V 4.10.4}. The data were corrected for the time-dependent gain problems following the standard technique. The event file was then screened to filter out strong background flares using 3$\sigma$ clipping algorithm \textit{lc\_clean} within CIAO, which yielded a net exposure time of 19.4 ks. Point sources within the frame were identified and removed using tool \textit{wavdetect}. The holes left behind after removal of point sources were re-filled for the imaging purpose by interpolating surrounding background counts using tool \textit{dmfilth}. As X-ray emission from this cluster covers major portion of the CCD chip, therefore, blank-sky frames provided by \textit{Chandra} X-ray Centre (CXC) were used for the background correction. Blank-sky frames thus acquired were processed in the same manner as for A2566, were normalized to the count rates of the source image in 9-12 keV energy band and then were subtracted as background from the source file. To account for the exposure variation of the data, the resultant background subtracted science frame was then divided by the exposure map generated using tool \textit{fluximage}. 
%%%%%%%%%%%%%%%%%%%%%%%%%%%%%%%%%%%%%%%%%%%%%%%%%%%%%%%%%%%%%%
\begin{figure*}
\centering
\includegraphics[scale=0.44]{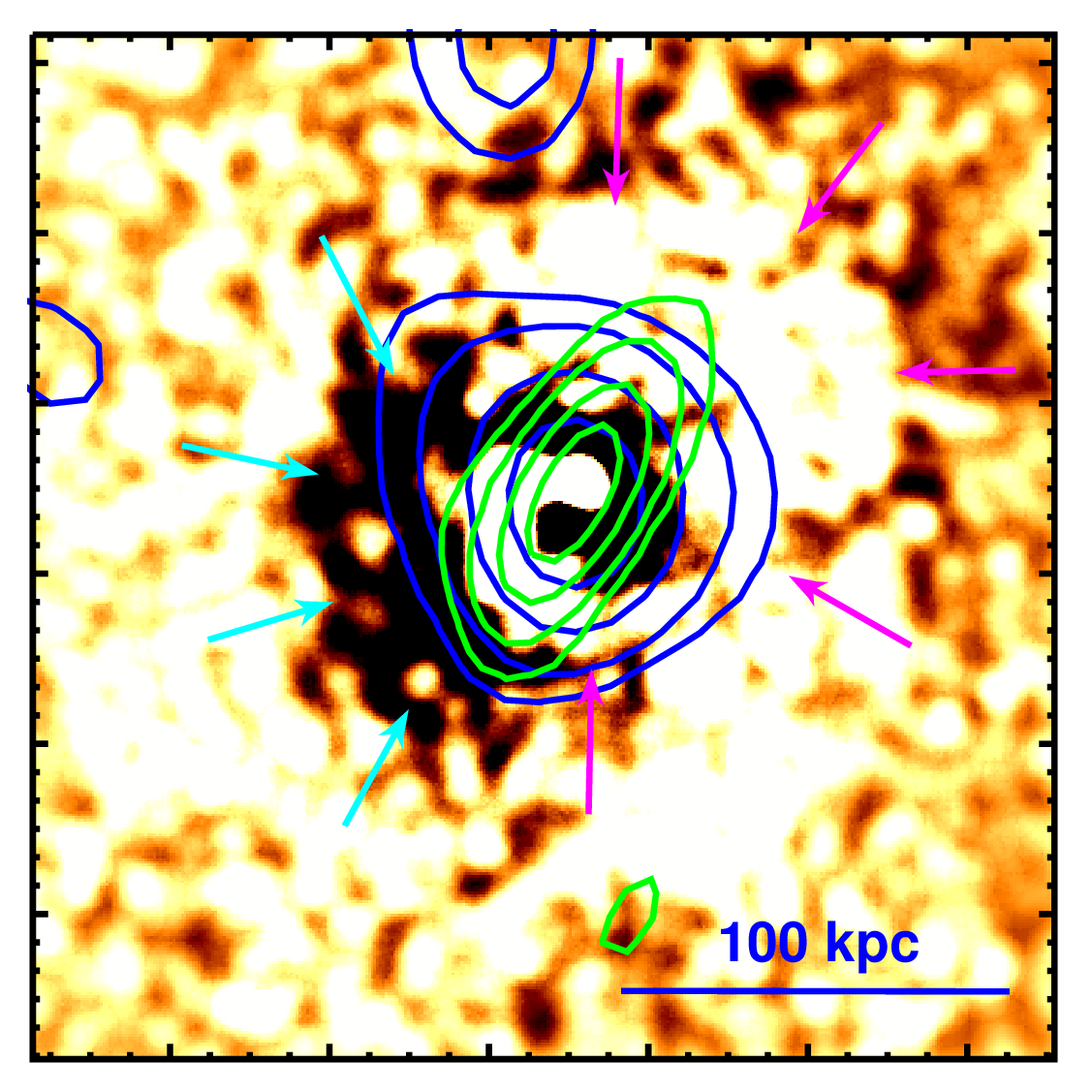} 
\includegraphics[scale=0.39]{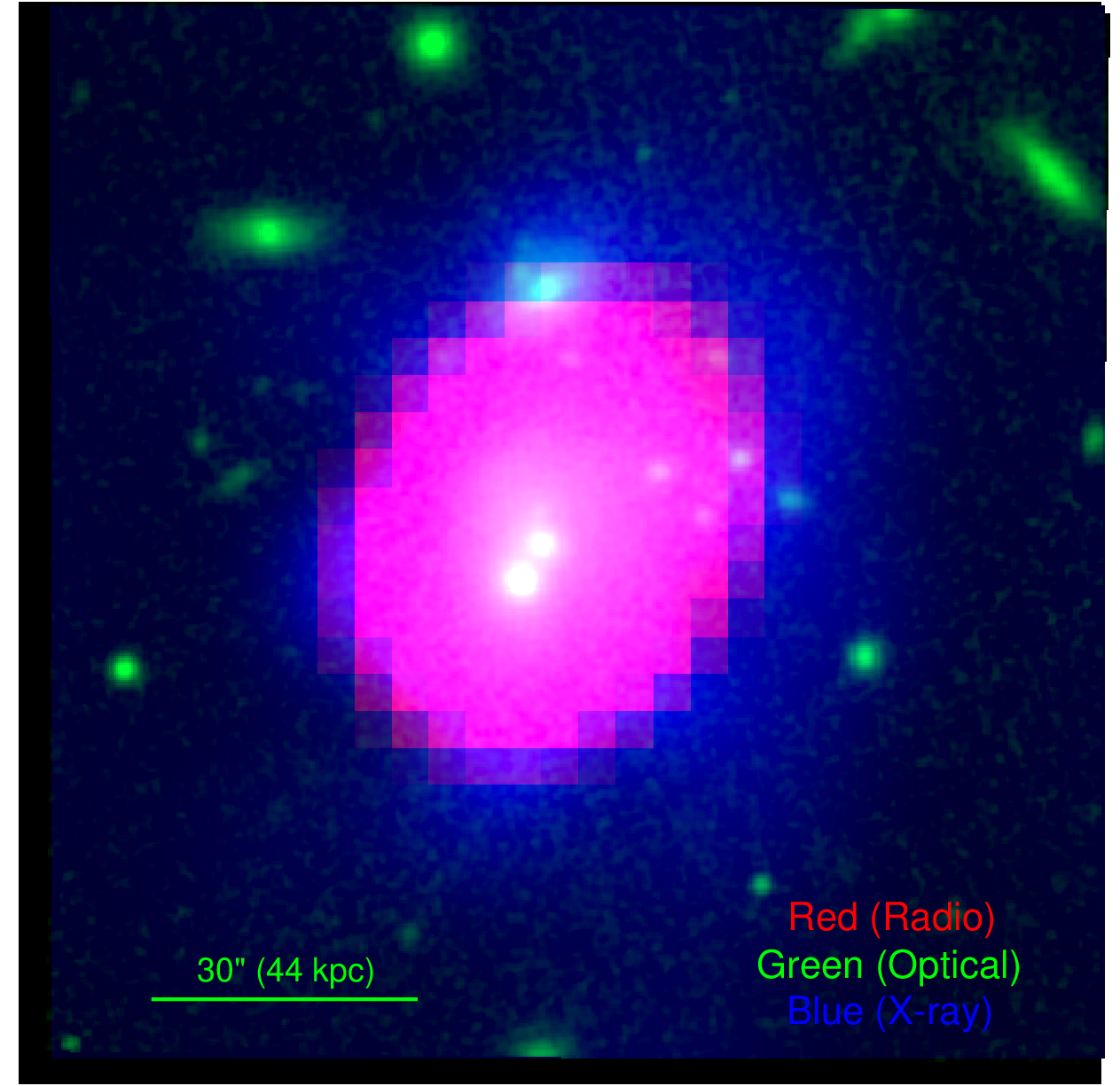}  
\caption[Residual Image and Composite Image]{\textit{Left panel:} 2D smooth model subtracted residual map of 0.5-3 keV cluster emission from A2566. For better visibility this image has been smoothed by a 10$\sigma$ Gaussian kernel. The green contours overlaid on it represent the radio emission detected at 1.4 GHz by VLA, while blue contours are due to the 150 MHz emission mapped using TGSS. Arrows in this figure highlight the position of spiral arms. \textit{Right panel:} RGB (tricolor) image of A2566 obtained by proper combination of emission measured at 1.4 GHz with VLA C-configuration (red), Pan-STARRS r-band (green), and \chandra X-ray soft band (blue).}
\label{resid_tri}
\end{figure*}
%%%%%%%%%%%%%%%%%%%%%%%%%%%%%%%%%%%%%%%%%%%%%%%%%%%%%%%%%%%%%%%%%%
%%%%%%%%%%%%%%%%%%%%%%%%%%%%%%%%%%%%%%%%%%%%%%%%%%%%%%%%%%%%%%%%%%
\begin{figure}
\includegraphics[scale=0.7]{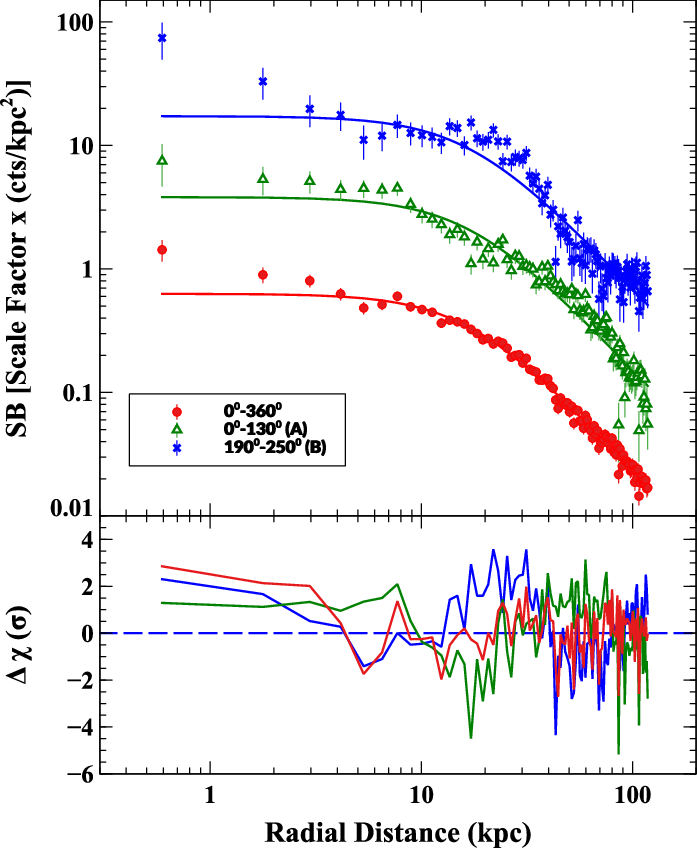}
\caption[Surface brightness profile]{Azimuthally averaged surface brightness profile of the X-ray emission from A2566 (red circles) along with the best fit $\beta$-model (continuous line). In the same figure we also plot the surface brightness profiles of the X-ray emission extracted from two wedge shaped regions covering 0\degr-130\degr (sector A) and 190\degr- 250\degr (sector B). For better presentation profiles along the sectors A and B are shifted along the ordinate axis by adding offset. Lower panel present deviations of the data relative to the best-fit models.}
\label{surface_brightness}
\end{figure}
%%%%%%%%%%%%%%%%%%%%%%%%%%%%%%%%%%%%%%%%%%%%%%%%%%%%%%%%%%%%%%%%%%

\section{Results}
\label{sec3}
\subsection{Imaging Analysis}
The exposure-corrected and background-subtracted 0.5-3 keV \textit{Chandra} X-ray image of A2566 (Figure~\ref{raw} left panel) reveals an asymmetric morphology of plasma distribution relative to the X-ray peak of the cluster emission at (RA, DEC) 23h16m04.71s and -20d27m46.70s. Optical counterpart of the X-ray emission from this cluster is shown in right panel of this figure and is the $r$-band image acquired using Pan-STARRS. This image clearly reveals interaction of the brightest cluster member ``BCG'' (NVSS J231604-202748; 2MASS J23160504-2027485) with a neighboring system ``S'' (SDSS J231605.02-202748.4). The two objects have a physical separation of 4.7 arcsec ($\approx$ 7 kpc), while the X-ray peak of the cluster (denoted by cross mark) is about 4.6 arcsec ($\approx$ 6.8 kpc) north-west of the BCG. Photometric redshift of the source S is $\sim$0.062581 \citep{2022ApJS..259...35A}, while that of the BCG is 0.08247, implying that the source ``S'' is indeed a cluster member.

To investigate hidden features within the cluster emission of A2566 we obtained its residual map by subtracting a best-fit two-dimensional beta model to the cluster surface brightness distribution. The 2D beta model was computed by fitting ellipses to the isophotes in the exposure-corrected, background-subtracted, point source removed 0.5-3 keV \textit{Chandra} image of A2566. The resultant residual map of A2566 shown in  Figure~\ref{resid_tri} (left panel) clearly reveals unusual distribution of the X-ray photons in this cluster delineating spiral-like structures. One of them exhibits spiral structure with excess emission (brighter shades) relative to the best-fit model, while its counterpart is a deficit region (darker shade) extended almost up to $\sim$75 kpc. For better representation this image has been smoothed by a 10$\sigma$ wide Gaussian kernel. To examine its association with the radio counterpart, we overlay 1.4 GHz Very Large Array (VLA) radio contours  (green) at levels 0.57, 1.91, 5.93, 12.63, 22 mJy (rms $\sim$ 190 $\mu$Jy, resolution $\sim$ 21.5 arcsec) and 150 MHz TGSS (blue) at levels 10.59, 19.67, 46.90, 92.29, 155.82 mJy (rms $\sim$ 3.5 mJy) on the same  map. The radio emission appears to be dispersed along the north-west to south-east direction and complements effectively the inner region of the spiral structure. Further, to examine resemblance of the spiral features at other wavelengths we have created a composite RGB image of A2566 (Figure~\ref{resid_tri} right panel) using emission in the optical, radio, and X-ray bands. In the background of this image is the optical emission mapped using Pan-STARRS $r$-band (shown in green), radio emission from the VLA at 1.4 GHz (red), and 0.5 to 3 keV \textit{Chandra} X-ray image tracing diffuse gas (blue). This figure clearly reveals that the diffuse emission in X-ray, optical, and radio band collectively follow identical orientation along the north-west to south-east of the BCG.

\subsection{X-ray Surface Brightness Profile}
\label{sb}
An azimuthally averaged surface brightness profile of the extended X-ray emission from A2566 was derived by extracting 0.5-3 keV counts from a series of concentric elliptical annuli centered on its X-ray peak extending up to 109 kpc. Resulting surface brightness profile was then fitted with an isothermal one dimensional $\beta$-model \citep{1976A&A....49..137C} assuming that the gas in the cluster is in hydrostatic equilibrium and was expressed as
\begin{equation}
S(r) = S_0 \left[ 1 + \left( \frac{r}{r_c} \right)^2 \right]^{- 3 \beta + 0.5}
\end{equation}
where, $S(r)$ represent surface brightness at a projected distance $r$, $S_0$ the central brightness, $\beta$ the slope parameter defined as ratio of the energy per unit mass contained in galaxies to that in the ICM, and $r_c$ the core radius. The azimuthally averaged surface brightness profile of the X-ray emission within this cluster is shown in Figure~\ref{surface_brightness} (red filled circles), while the best fit model is represented by the red continuous line. The best fit model yielded the slope parameter $\beta$ equal to $0.47\pm0.02$, while the core radius was found to be $17.2\pm0.40$ kpc, in good agreement with those reported earlier for cool core galaxy clusters \citep{2012MNRAS.421..808P, 2015Ap&SS.359...61S, 2016MNRAS.461.1885V, 2019MNRAS.484.4113K}.

Residual map of the X-ray emission from A2566 revealed several fluctuations including spiral structures formed due to the gas sloshing. To examine these features in further details we have also computed surface brightness profiles of the X-ray emission from A2566 by extracting 0.5-3 keV counts from two wedge shaped sectorial regions shown in Figure~\ref{raw} (left panel) with angular coverage of 0\degr-130\degr (sector A) and 190\degr-250\degr (sector B), both measured from the west to north. The sectorial surface brightness profiles and the best fit models are shown in Figure~\ref{surface_brightness}, respectively, by open triangles (green) and asterisks (blue). For better visibility these sectorial profiles were shifted along the ordinate axis by adding offsets. While surface brightness along sector A exhibit emission deficiency between 10-30 kpc due to the apparent darker spiral structure in residual map, significant excess emission corresponding to its brighter counterpart is evident between 10 to 40 kpc in the profile for sector B. These profiles also show edges in the emission at 77.4 kpc at 41.6 kpc, respectively, along sectors A and B. The edge along sector A is relatively shallow compared to that along B. We also investigate thermodynamical properties of the ICM along these wedge shaped sectors by performing spectral analysis X-ray photons and fitting deprojected broken power-law density function to it and are discussed in Section~\ref{PROFFIT}.

%%%%%%%%%%%%%%%%%%%%%%%%%%%%%%%%%%%%%%%%%%%%%%%%%%%%%%%%%%%%%%%%%%%%%%%%%%%%%
\begin{figure*}
	\centering
	{
			\includegraphics[scale=0.60]{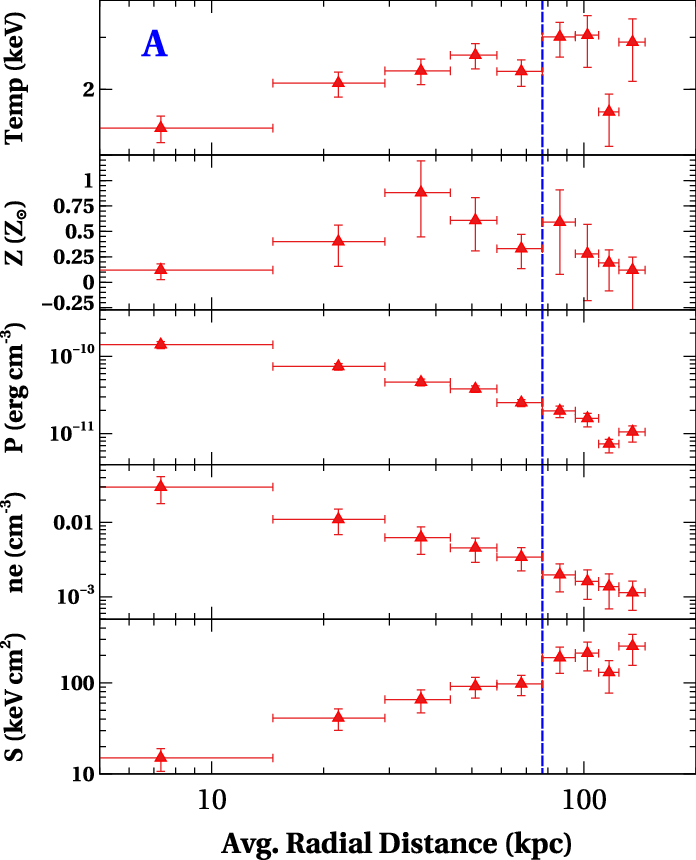}
   \hspace{0.5cm}
		\includegraphics[scale=0.61]{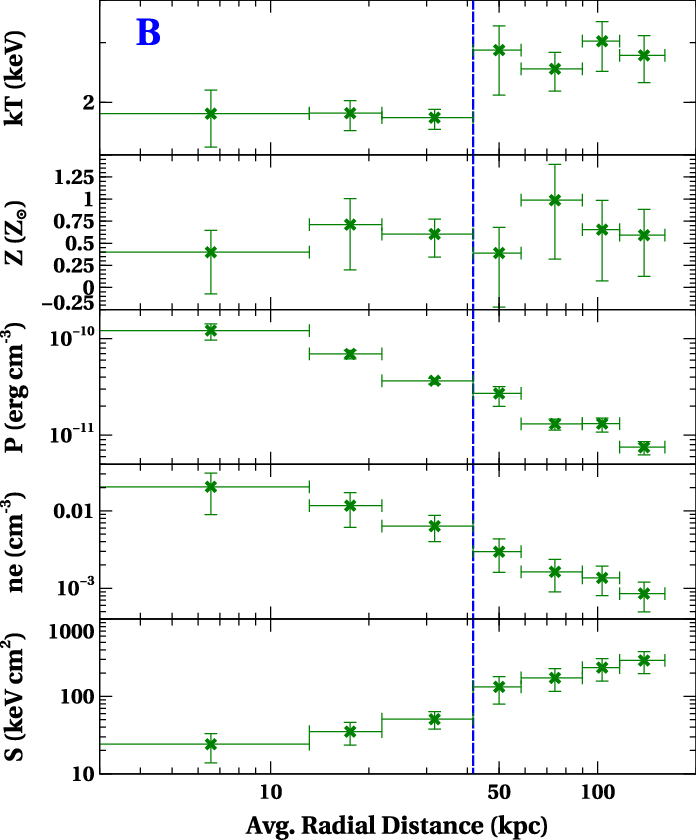}
	}
	\caption{Radial profiles of the thermodynamical properties of the ICM like, temperature ($kT$), metallicity ($Z$), gas pressure ($p$), electron density ($n_e$) and entropy ($S$) along sector A (\textit{left panel}) and sector B (\textit{right panel}), respectively. Blue dashed vertical lines in both the plots correspond to the position of cold fronts.} 
	\label{azmuth_spec}
\end{figure*} 
%%%%%%%%%%%%%%%%%%%%%%%%%%%%%%%%%%%%%%%%%%%%%%%%%%%%%%%%%%%%%%%%%%%%%%%%%%%%%%

%%%%%%%%%%%%%%%%%%%%%%%%%%%%%%%%%%%%%%%%%%%%%%%%%%%%%%%%%%%%%%%%%%

\begin{figure*}
\centering
\includegraphics[scale=0.43]{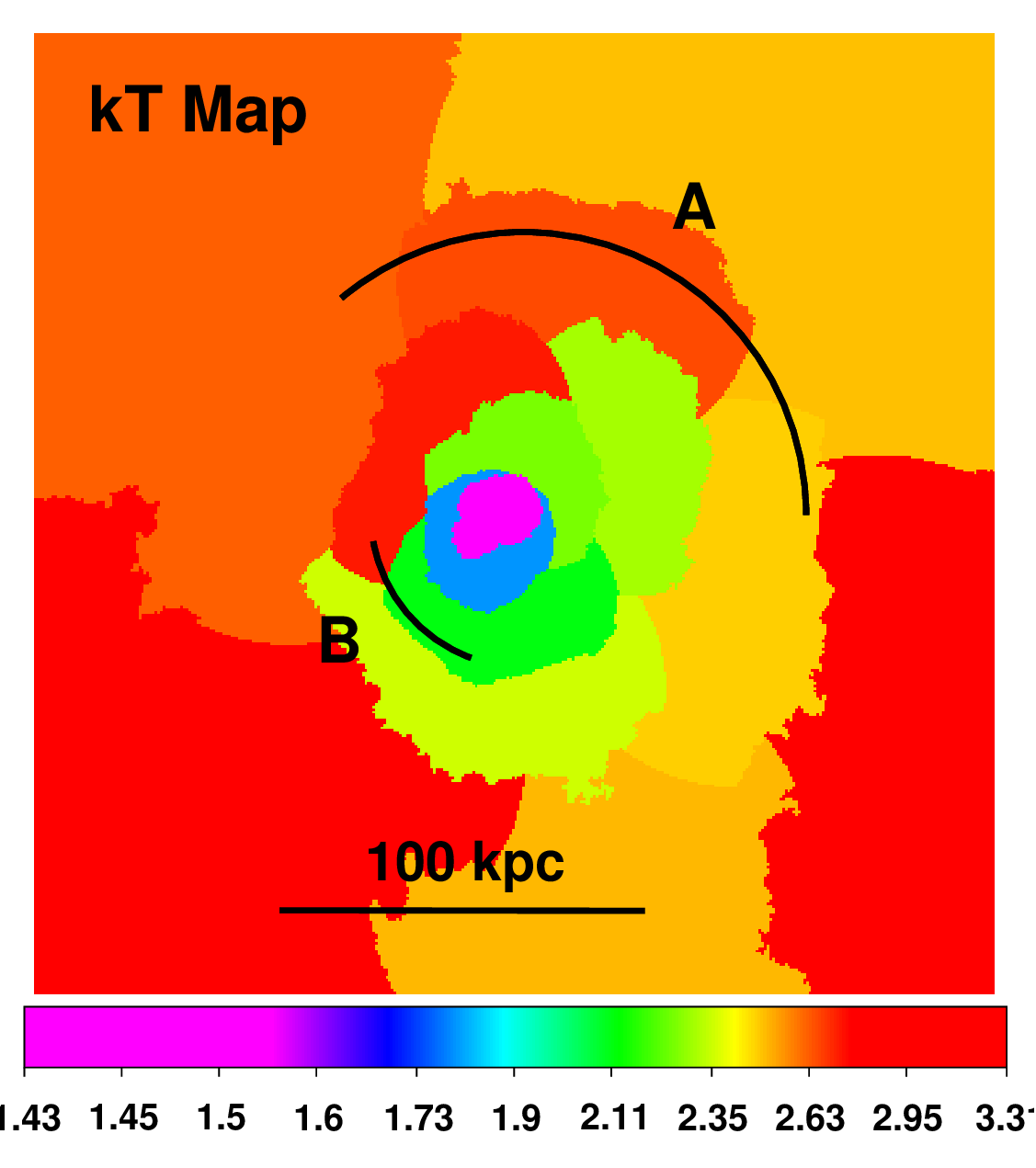}
\hspace{0.5cm}
\includegraphics[scale=0.43]{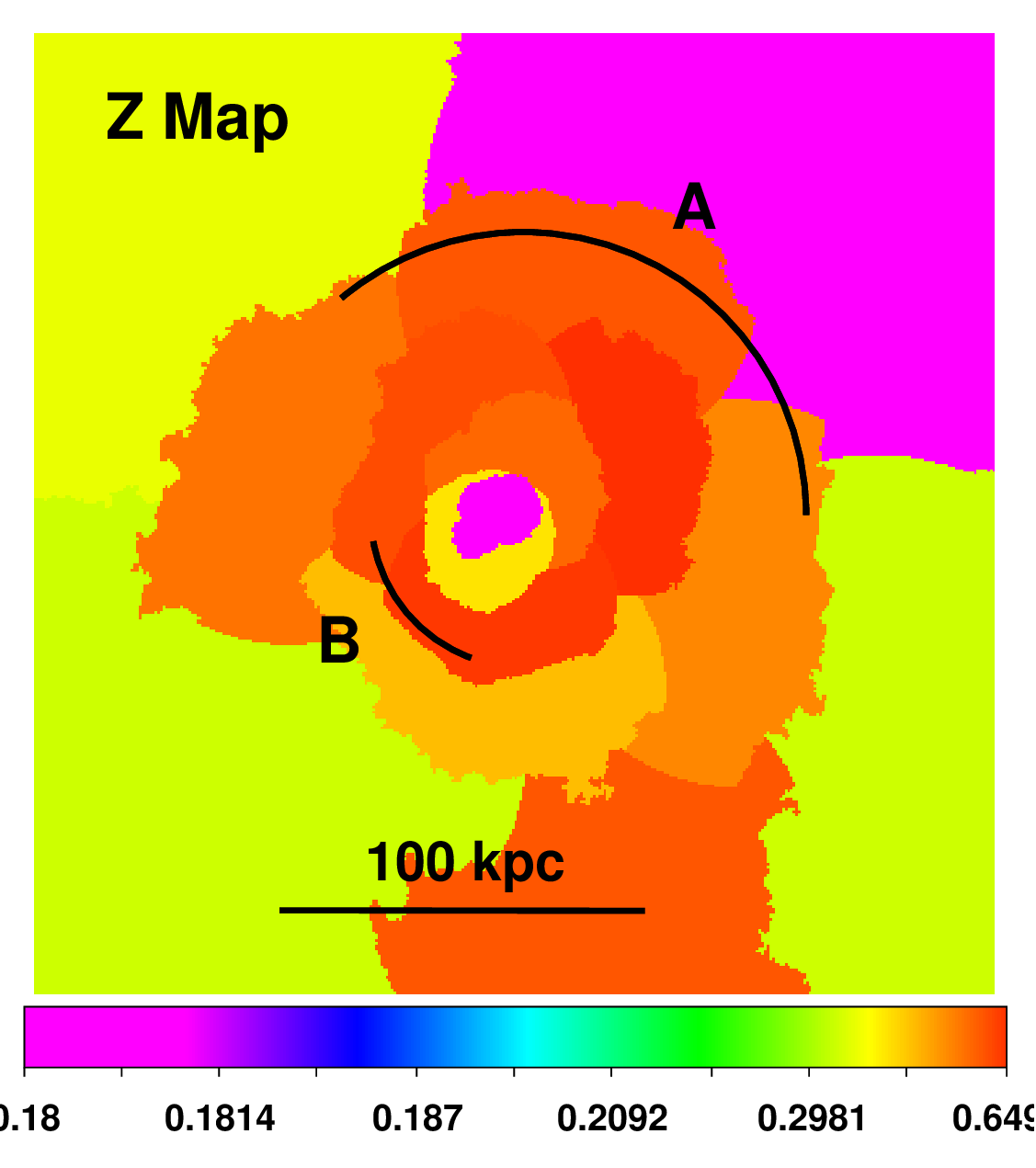}
\caption[T Z map]{2D temperature map in units of keV (\textit{left panel}) and metallicity map in units of Z$_{\odot}$ (\textit{right panel}). Cold fronts along sectors A and B are highlighted by the black arcs.}
\label{kT_Z_map}
\end{figure*}
%%%%%%%%%%%%%%%%%%%%%%%%%%%%%%%%%%%%%%%%%%%%%%%%%%%%%%%%%%%%%%%%%%

\subsection{Spectral Analysis} 
\label{spectra_ana}
To examine the radial dependence of thermodynamical properties of the ICM along wedge shaped sectorial regions A and B, we extract 0.5 - 7 keV X-ray spectra from the cleaned event file using script \textit{specextract} within CIAO \citep[for details please refer to][]{2017MNRAS.472.2042P}. Widths of the sectorial bins were set such that each extraction yielded a minimum of 500 background subtracted counts. For the background treatment we extracted spectra from the blank-sky corresponding to the source regions. Extracted spectra from all the sectorial bins were then fitted with a single temperature \textsc{apec} model within \textsc{xspec v 12.12.0} \citep{1996ASPC..101...17A}. To apply $\chi^2$ statistics to the data, spectra were binned to a minimum of 20 counts. The Galactic absorption correction was applied following \textit{tbabs} model of \cite{2000ApJ...542..914W} and fixing the column density at $\rm N_H$ at $\rm 1.90\times 10^{20}\, cm^{-2}$ \citep{2005yCat.8076....0K}. Other thermodynamical quantities like, temperature, abundance and normalization were allowed to vary during the fit. The resultant projected profiles of the thermodynamical quantities for both the sectorial regions are shown in Figure~\ref{azmuth_spec}.

We also derive other properties of the ICM such as electron density, pressure, and entropy for each of the extraction. Here, the electron density was computed using the {\sc apec} normalization obtained above and following the expression in \cite{2022ApJ...938...51A}

\begin{equation}
    n_e  = \biggl[1.2 \ N \times 4.07\times10^{-10}(1+z)^2 \\ \biggl (\frac{D_A}{\rm{Mpc}}\biggl)^2\biggl(\frac{V}{\rm{Mpc^3}}\biggl)^{-1}\biggl]^{1/2},
\label{eq:density}
\end{equation}

Here, $D_A$ represent the angular diameter distance of the source, $V$ volume of the extraction region, $N$ the {\sc apec} normalization. We assume $n_e/n_H \sim$1.18 \citep{1989A&A...215..147B}; with $n_e$, $n_H$, respectively, being the electron and hydrogen densities. We then derive ICM pressure and the entropy for each of the region using $p=nkT$ and $S=kT/n_e^{2/3}$ \citep{2003MNRAS.343..331P}, $n$ being the gas density ($n\sim 1.92 n_e$). The resultant projected radial profiles of the ICM pressure and the entropy along sectors A and B are shown in Figure~\ref{azmuth_spec}. The entropy profiles derived along both the sectors exhibit systematic rise as a function of radial distance. The larger error bars in the metallicity profiles along these sectorial regions do not allow us to arrive at proper conclusion, however, appears to be consistent with the average value of $0.54\pm0.18$ $Z_{\odot}$.

\subsection{2D Temperature and Metallicity Map}
With an objective to examine spatial variations in the temperature and metallicity of the ICM within cluster environment we derive 2D maps of temperature and metallicity following the adaptive binning technique of \cite{2006MNRAS.371..829S}. For this we perform spatially resolved spectroscopic analysis of the X-ray photons extracted from the rectangular regions produced by the \textit{contbin} algorithm. This method identifies the brightest pixels in the point source removed X-ray image and generate spatial bins around such pixels covering their neighbors of similar brightness. This process was then repeated till it reaches the predefined S/N of $\sim$35. Spectra for each of the detected region were then extracted and treated in the same way as discussed above (Section~\ref{spectra_ana}). This in conjunction with the \textit{paint\_output\_images} task yielded the 2D maps of the temperature and metallicity and are shown in Figure~\ref{kT_Z_map}. The errors in the temperature map corresponds to $\sim$6\% in the inner region and $\sim$9\% in the outer region, while those in the abundance map ranges from 18\% to 25\%. The temperature map shows a noticeable jumps apparent in arc shapes on the east as well as on the west of the X-ray peak, which coincides with the locations of cold fronts. Though similar trends are also evident in the metallicity map, however, large error bars do not permit us to constrain it. 

%%%%%%%%%%%%%%%%%%%%%%%%%%%%%%%%
\begin{figure*}
\centering
\includegraphics[scale=0.45]{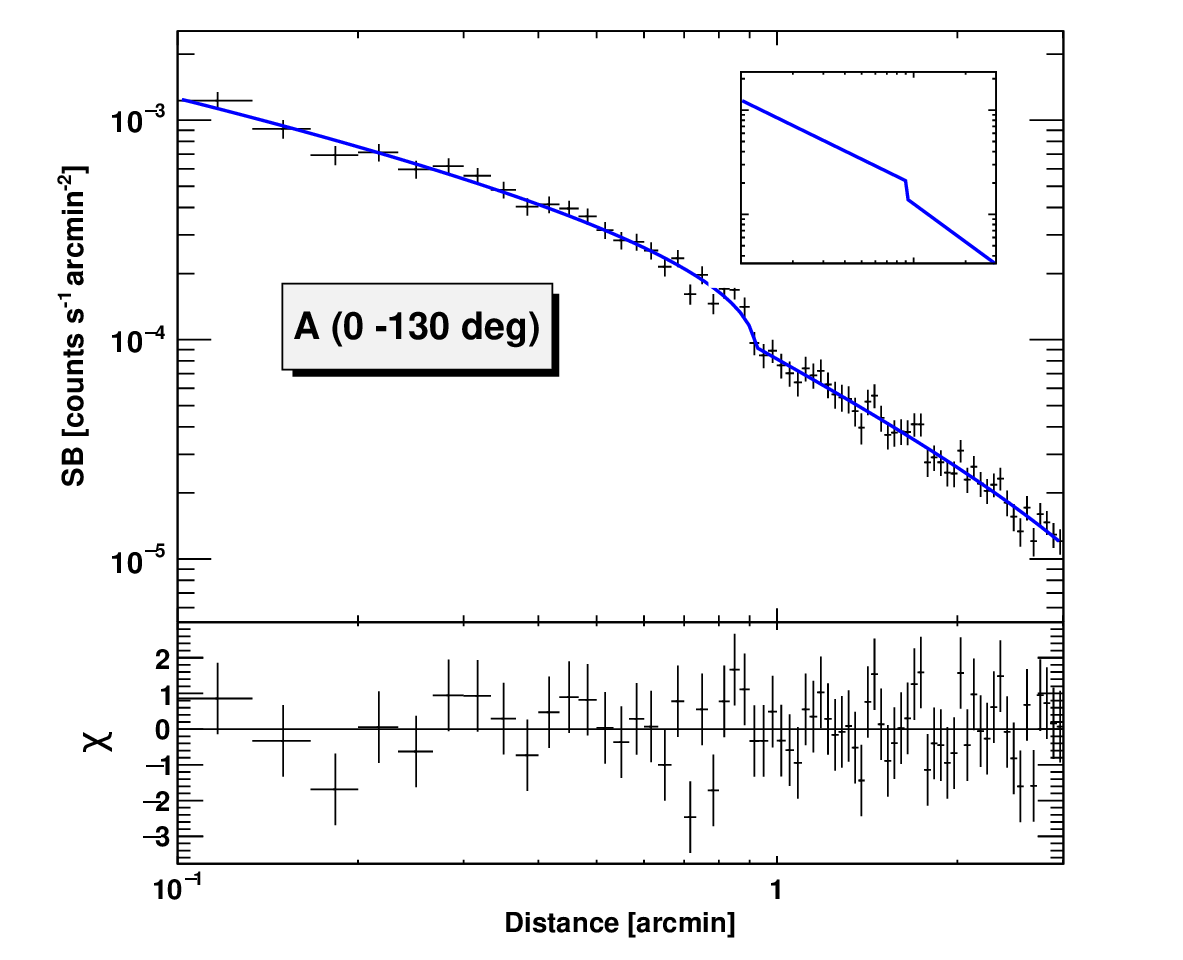}
%\hspace{-1.1cm}
\includegraphics[scale=0.45]{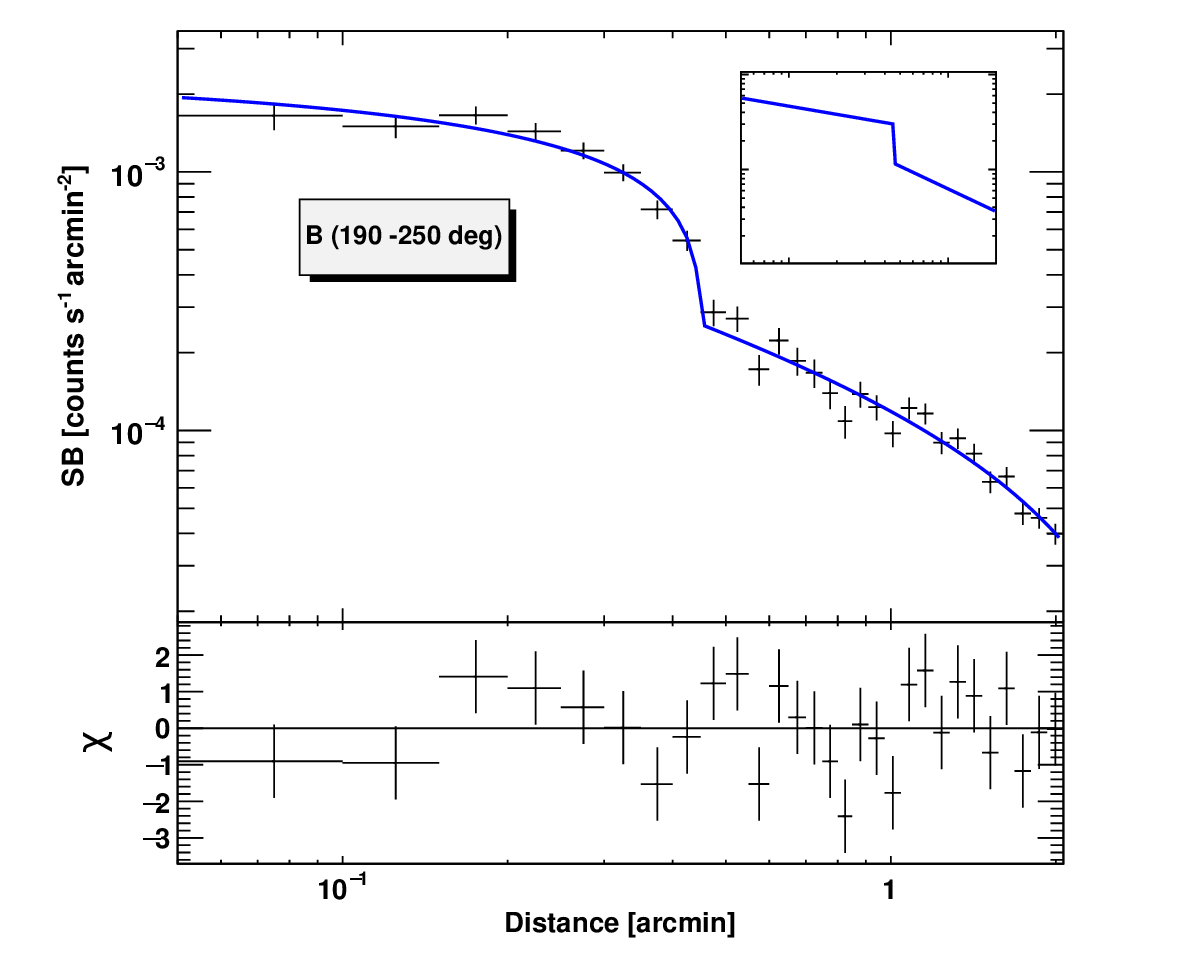}
\caption{Broken power-law density models fitted to the 0.5–3 keV X-ray surface brightness along the wedge shaped sectors A {(\it left panel)} and B {(\it right panel)}. The error bars shown are at the 1$\sigma$ level. The best-fit deprojected density profile model are shown in the insets.}
\label{proffit}
\end{figure*}
%%%%%%%%%%%%%%%%%%%%%%%%%%%%%%%%%%%%%%%%%%%%%%%%%%%%%%%%%%%%%%%%%%%%%%%%%%%%%%%%%%%%%%%%%%%%%%%%%
%%%%%%%%%%%%%%%%%%%%%%%
\begin{table*}
\centering
\caption{Best fit parameters of the broken power-law density function to the surface brightness}
\begin{tabular}{|c|c|c|c|c|c|c|}\hline
Regions              &$\alpha$1&     $\alpha$2  &$r_{\text{edge}}$    &$n_0$       &C               &$\chi^{2}$/dof  \\
&&&(arcmin)&($10^{-4}\rm\,cm^{-3}$)&&     \\ \hline
A (0\degr-130\degr) &$0.81\pm0.03$ &$1.21\pm0.11$ &$0.91\pm0.01$ &$1.92\pm0.10$ &$1.48\pm0.09$  &54.75/63     \\

B (190\degr-250\degr)&$0.25\pm0.08$ &$0.81\pm0.12$ &$0.44\pm0.01$ &$25\pm2.50$ &$2.59\pm0.22$  &33.61/23     \\  \hline
 
\end{tabular}
\label{tab585}
\end{table*}
% %%%%%%%%%%%%%%%%%%%%%%%%%%%%%%%%%%%%%%%%%%%%%%%%%%%%%%%%%%%%%%%%%%%%%%%%%%%%%%%%%%%%%%%%%%%%%%%%%%%%

\section{Discussion}
\label{disc}
\subsection{Detection of Cold Fronts}
\label{PROFFIT}
As discussed above, the surface brightness profiles along the sectorial regions A and B revealed edges in the plasma distribution and are likely associated with the cold fronts.  To confirm such an association we fit deprojected broken power-law density function of \cite{2011A&A...526A..79E} to the surface brightness distribution along both the wedge shaped sectorial regions. This was achieved by using interactive package PROFFIT (version 1.4), which provides a potential way to identify discontinuity in the surface brightness \citep{2009ApJ...704.1349O} as

	\begin{equation}
		n_e(r) = {\left\{\begin{matrix}
				C  n_0 \left(\frac{r}{r_{\text{edge}}}\right)^{-\alpha_1} \quad                      \text{ if } r \leq r_{\text{edge}} \\ 
				n_0 \left(\frac{r}{r_{\text{edge}}}\right)^{-\alpha_2} \quad \text{ if } r > r_{\text{edge}}
			\end{matrix}\right.}
	\end{equation}
	
here, $n_e(r)$ represent the electron number density at a projected distance $r$ from the cluster peak, $r_{edge}$ location of the putative edge, $C$ represent the density compression factor, $\alpha_1$ and $\alpha_2$, respectively, as the inner and outer slopes of the power law, while $n_0$ represent the density normalization. The underlying assumption of the model is that the ICM is in hydrostatic equilibrium and is spread in symmetric relative to its core. During the fit we allowed $\alpha_1$, $\alpha_2$, $C$, and $n_0$ to vary. 
	
The resultant surface brightness profile fitted with the broken power-law is shown in Figure~\ref{proffit} and the best-fit quantities are listed in Table~\ref{tab585}. The profile along sector A clearly exhibits sharp density break at $\sim$77.4 kpc due to its association with the cold front, while that along sector B reveals a similar break at $\sim$41.6 kpc. Thus, the edges in surface brightness profiles along both the sectors A and B were confirmed to be associated with the cold fronts. Spectral analysis of the X-ray photons along sector A also revealed a jump in the temperature from $2.3^{+0.25}_{-0.21}$ keV to $3.1^{+0.56}_{-0.46}$ keV at the location of cold front, while pressure remained almost constant across the discontinuity ($2.5^{+0.27}_{-0.23}\times10^{-11}$ erg\, cm$^{-3}$ and $1.9^{+0.35}_{-0.29}\times10^{-11}$ erg\, cm$^{-3}$). A sharp drop in the gas density almost by a factor of 1.48 was also evident at this location. On the other side, the cold front along sector B exhibited a temperature rises from $1.8^{+0.10}_{-0.07}$ keV to $2.8^{+0.74}_{-0.50 }$ keV with a continuity in the pressure profile and a drop in the gas density almost by a factor of 2.6. Such abrupt changes in the quantities across the discontinuity along sectors A and B are indeed due to the presence of cold fronts and are consistent with the findings of several other researchers e.g., \cite{2022A&A...664A.186S, gopal-krishna_paul_salunkhe_sonkamble_2022}.
	
\subsection{Origin of the Sloshing and the Cold Fronts}
\label{disc2}

2D imaging analysis of \textit{Chandra} X-ray data on A2566 revealed unusual morphology of ICM distribution that appeared in the form of spiral-shaped gas sloshing along with edges in the surface brightness distribution. Spectral analysis of the X-ray photons from the two sectorial regions revealed association of these morphological discontinuities with the cold fronts. Further, this study also witnessed close association of the BCG with a neighboring system, off-set in the X-ray peak of the cluster emission and its close association with the diffuse radio emission emanating from the core of this cluster. All these features collectively together make A2566 an interesting system and point towards common origin of all these features from a plausible minor merger like episode. The gas sloshing due to an interaction with a cluster member in A2566 were already pointed by \cite{2016MNRAS.460.1758H}. Another clue for such an interaction was provided by the offset in the H$\alpha$ \citep{2012MNRAS.421.3409H}. All the observed features along with offset in the H$\alpha$ emission peak are in agreement with those reported in \citep[RXJ1347.5-1145,][]{2012ApJ...751...95J}, which were suggested to be originated due to a minor merger. The complex morphology of plasma distribution in A2566 is also likely due to a minor merger of a sub-cluster that disturbed the main cluster by  displacing gravitational potential well of it. Such a displacement further results in the formation of cold fronts, the concentrically shaped borders in the surface brightness produced by the core's gas as it moves around the potential well. These cold fronts further develop spiral patterns in the plasma distribution provided the sloshing direction is close to the plane of sky \citep{2006ApJ...650..102A,2007PhR...443....1M}. In the case of A2566 spectral analysis of the X-ray photons extracted along the two wedge shaped sectors A and B revealed discontinuities in the temperature profiles at the location of cold fronts while maintaining pressure balance across it (Figure~\ref{azmuth_spec}). The observed evidences of spiral structure in the plasma distribution, interaction of sub-cluster S with the BCG, cold fronts associated with the spiral structure, all collectively suggest that they have common origin due to minor merging. 

\subsection{Association of X-ray and H$\alpha$ emission}
\label{disc3}
\begin{figure}
\includegraphics[scale=0.44]{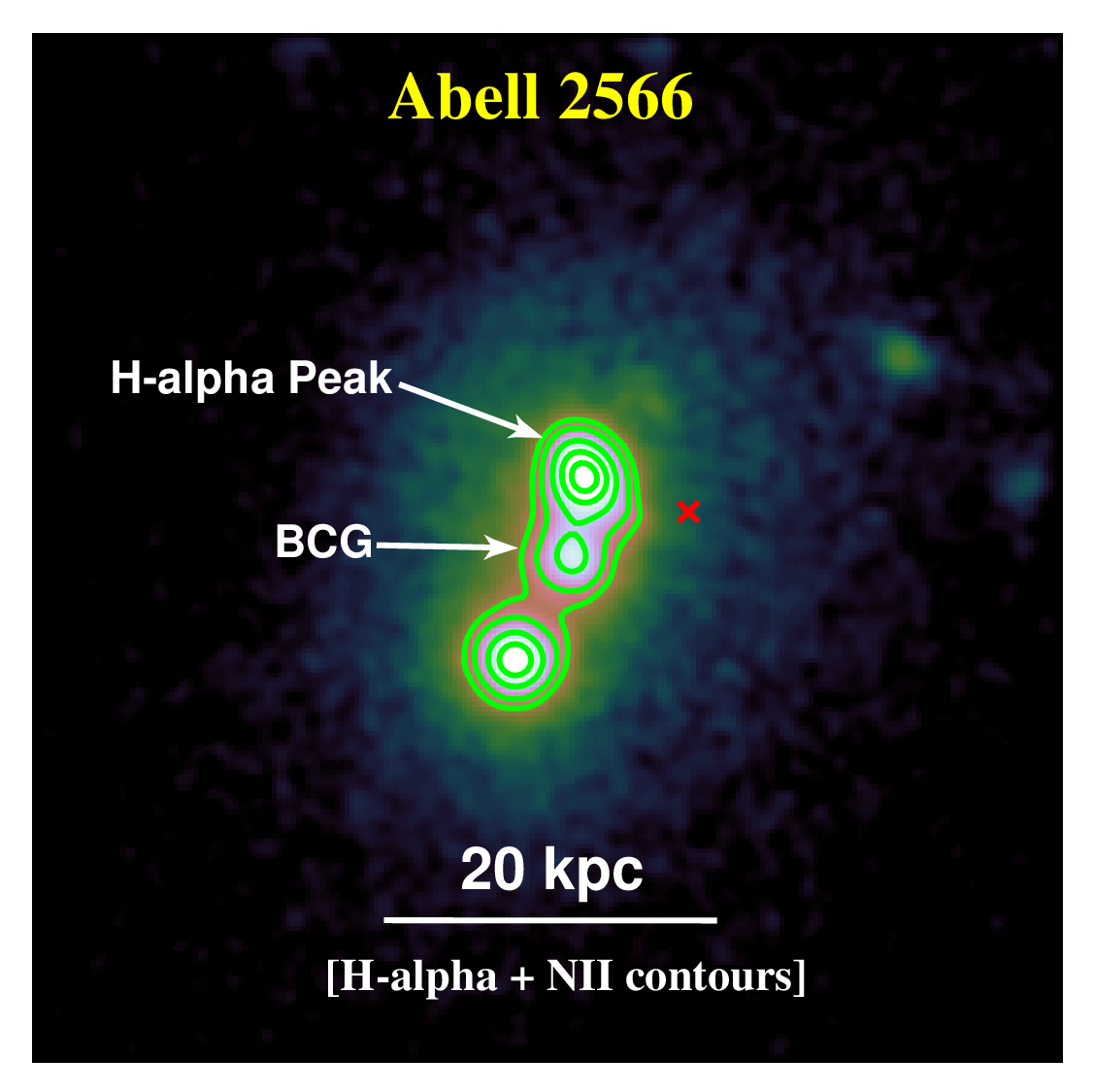}
\caption[T Z map]{MUSE H$\alpha$+[NII] emission map with contours. Red cross in this figure denotes the location of X-ray emission peak of A2566.}
\label{muse}
\end{figure}

Azimuthally averaged 0.5 - 7 keV temperature profile showed that the brightest core of the cluster hosts cool ICM, demonstrating that at the core of A2566 lies a rapidly cooling gas. This core of cool gas appears to be the primary source of H$\alpha$ emission and not the BCG, challenging the conventional belief that H$\alpha$ emission primarily originate from the BCG. This in turn highlights the importance of understanding the dynamics of cooling gas in the core of this cluster. \cite{2016MNRAS.460.1758H} during their study have confirmed an offset between the positions of the H$\alpha$ emission peak and the X-ray peak. Such an offset has also been evidenced in Figure~\ref{muse}, where the X-ray peak is denoted by the cross. This image was produced using the Multi-Unit Spectroscopic Explorer (MUSE) - Integral Field Unit (IFU) observations acquired with the Very Large Telescope (VLT). The existence of similar offset between H$\alpha$ and X-ray peaks have also been reported in several other clusters like, Abell 1644 \citep{2010ApJ...710.1776J}, RXJ 1347.5-1145 \citep{2011PhDT........14J}, Abell 1991 \citep{2013Ap&SS.345..183P}, Abell 1668 \citep{2021ApJ...911...66P}, and the Ophiuchus cluster \citep{2012MNRAS.421.3409H}. Such consistent offsets across multiple clusters point toward a potential association between the H$\alpha$ emitting molecular gas and plasma in the deepest potential well. According to \cite{2011MNRAS.417.3080C} the observed offsets are due to the merging events that separates BCG and the H$\alpha$ emitting gas from the X-ray emission peak. Thus, the merger induced sloshing processes might be responsible for the transient dissociation of the X-ray peak from the BCG and H$\alpha$. Convincing similarity in the sloshing structures and offset characteristics between A2566 and other clusters like, Abell 1644 and RXJ 1347.5-1145 strengthens the assumption of their common origin.

%%%%%%%%%%%%%%%%%%%%%%%%%%%%%%%%%%%%%%%%%%%%%%%%%%%%%%%%%%%%%%%%%%
%%%%%%%%%%%%%%%%%%%%%%%%%%%%%%%%%%%%%%%%%%%%%%%%%%%%%%%%%%%%%%%%%%
\subsection{Spatial association of X-ray peak with the BCG}
For the case of dynamically relaxed galaxy clusters it is believed that the BCG would be at the center of the gravitational potential \citep{2005MNRAS.361.1203V}. Therefore, peak of the X-ray emission from such clusters can act as a robust and reliable diagnostic indicator for assessing its dynamical state, whether they are relaxed or not. However, several of the X-ray emission studies have revealed that the X-ray peaks in many of the clusters exhibit offsets relative to the position of the BCGs, pointing toward their disturbed states \citep[see][]{2012MNRAS.420.2120M,2016MNRAS.457.4515R,2021MNRAS.505.2628P}. This study also reveals an offset of about 4.6\arcsec\, (6.8 kpc) between the X-ray peak and the optical position of the BCG. Coordinates of the BCG are R.A. (J2000) 23h16m05.02s, DEC. (J2000) -20d27m48.23s and correspond to the brightest pixel of the optical Pan-STARRS (r band) image. Though offset is a prime indicator for disturbed morphology, recent statistical studies by \cite{2016MNRAS.457.4515R} demonstrated that the offsets less than 0.02 $R_{500}$ may be common even in the relaxed cluster. For the case of A2566 we compute $R_{500}$ to be 866.37 kpc based on the $M_{500}$ mass, critical density of the Universe, assuming cosmological parameters and following method described in \cite{2023MNRAS.526.1367J}. This implies that the offset of 6.8 kpc is equivalent to 0.008 $R_{500}$ and is in agreement with the fact that A2566 is a relatively relaxed cluster.

\section[4]{Conclusions}
\label{con}
This study presents analysis of X-ray data on apparently relaxed cluster A2566. 2D imaging analysis of the \textit{Chandra} data revealed considerable spiral structures in the morphology of X-ray emission from  within the central 109 kpc of this cluster. Additionally, two sharp edges on the south-east and north-west directions of the X-ray peak of the cluster at $\sim$41.6 kpc and $\sim$77.4 kpc, respectively, have also been evidenced in the surface brightness distribution. A detailed analysis of the sectorial brightness profiles along these edges confirm their origin due to sloshing of gas, referred to as the sloshing cold fronts. Spectral analysis of 0.5 - 7 keV X-ray photons along these discontinuities exhibited sharp temperature jumps from 2.3 to 3.1 keV and 1.8 to 2.8 keV along the north-west and south-east directions, respectively. Interestingly, continuity in the pressure profiles across these discontinuities point towards their association with the cold fronts. Further confirmation for such  associations were provided by the deprojected broken power-law density functions fitted to the surface brightness distribution along both these wedge shaped sectorial regions. It is hypothesized that offset between the X-ray emission peak and BCG might have yielded the sloshing structure and have originated by minor mergers.

\section*{Acknowledgments} 
{SKK gratefully acknowledges the financial support of UGC, New Delhi, under the Rajiv Gandhi National Fellowship (RGNF) scheme. MKP and NDV thanks to IUCAA for supporting the library facility. This study has made use of the data from \chandra Data Archive, NASA/IPAC Extragalactic Database (NED), and High Energy Astrophysics Science Archive Research Center (HEASARC). This work acknowledges the use of software packages CIAO and Sherpa provided by the \textit{Chandra} X-ray Center.

\def\nat{Nature}%
\def\aj{AJ}%
\def\actaa{Acta Astron.}% % Acta Astronomica
\def\araa{ARA\&A}% % Annual Review of Astron and Astrophys
\def\apj{ApJ}% % Astrophysical Journal
\def\apjl{ApJ}% % Astrophysical Journal, Letters
\def\apjs{ApJS}% % Astrophysical Journal, Supplement
\def\aap{A\&A}% % Astronomy and Astrophysiof
\def\aapr{A\&A~Rev.}% % Astronomy and Astrophysics Reviews
\def\aaps{A\&AS}% % Astronomy and Astrophysics, Supplement
\def\apss{Ap\&SS}% %Astrophysics and Space Science
\def\baas{BAAS}% % Bulletin of the AAS
\def\caa{Chinese Astron. Astrophys.}% % Chinese Astronomy and Astrophysics
\def\cjaa{Chinese J. Astron. Astrophys.}% % Chinese Journal of Astronomy and Astrophysics
\def\icarus{Icarus}% % Icarus
\def\jcap{J. Cosmology Astropart. Phys.}% % Journal of Cosmology and Astroparticle Physics
\def\jrasc{JRASC}% % Journal of the RAS of Canada
\def\memras{MmRAS}% % Memoirs of the RAS
\def\mnras{MNRAS}% % Monthly Notices of the RAS
\def\na{New A}% % New Astronomy
\def\nar{New A Rev.}% % New Astronomy Review
\def\pra{Phys.~Rev.~A}% % Physical Review A: General Physics
\def\prb{Phys.~Rev.~B}% % Physical Review B: Solid State
\def\prc{Phys.~Rev.~C}% % Physical Review C
\def\prd{Phys.~Rev.~D}% % Physical Review D
\def\pre{Phys.~Rev.~E}% % Physical Review E
\def\prl{Phys.~Rev.~Lett.}% % Physical Review Letters
\def\pasa{PASA}% % Publications of the Astron. Soc. of Australia
\def\pasp{PASP}% % Publications of the ASP
\def\pasj{PASJ}% 
\def\sovast{SOVAST}% 
\def\ssr{Space Sci. Rev.}
\def\physrep{Phys. Rep.}

\bibliographystyle{elsarticle-harv} 
\bibliography{mybib.bib}

\begin{thebibliography}{62}
\expandafter\ifx\csname natexlab\endcsname\relax\def\natexlab#1{#1}\fi
\providecommand{\url}[1]{\texttt{#1}}
\providecommand{\href}[2]{#2}
\providecommand{\path}[1]{#1}
\providecommand{\DOIprefix}{doi:}
\providecommand{\ArXivprefix}{arXiv:}
\providecommand{\URLprefix}{URL: }
\providecommand{\Pubmedprefix}{pmid:}
\providecommand{\doi}[1]{\href{http://dx.doi.org/#1}{\path{#1}}}
\providecommand{\Pubmed}[1]{\href{pmid:#1}{\path{#1}}}
\providecommand{\bibinfo}[2]{#2}
\ifx\xfnm\relax \def\xfnm[#1]{\unskip,\space#1}\fi
%Type = Article
\bibitem[{{Abdurro'uf} et~al.(2022){Abdurro'uf}, {Accetta}, {Aerts}, {Silva Aguirre}, {Ahumada}, {Ajgaonkar}, {Filiz Ak}, {Alam}, {Allende Prieto}, {Almeida}, {Anders}, {Anderson}, {Andrews}, {Anguiano}, {Aquino-Ort{\'\i}z}, {Arag{\'o}n-Salamanca}, {Argudo-Fern{\'a}ndez}, {Ata}, {Aubert}, {Avila-Reese}, {Badenes}, {Barb{\'a}}, {Barger}, {Barrera-Ballesteros}, {Beaton}, {Beers}, {Belfiore}, {Bender}, {Bernardi}, {Bershady}, {Beutler}, {Bidin}, {Bird}, {Bizyaev}, {Blanc}, {Blanton}, {Boardman}, {Bolton}, {Boquien}, {Borissova}, {Bovy}, {Brandt}, {Brown}, {Brownstein}, {Brusa}, {Buchner}, {Bundy}, {Burchett}, {Bureau}, {Burgasser}, {Cabang}, {Campbell}, {Cappellari}, {Carlberg}, {Wanderley}, {Carrera}, {Cash}, {Chen}, {Chen}, {Cherinka}, {Chiappini}, {Choi}, {Chojnowski}, {Chung}, {Clerc}, {Cohen}, {Comerford}, {Comparat}, {da Costa}, {Covey}, {Crane}, {Cruz-Gonzalez}, {Culhane}, {Cunha}, {Dai}, {Damke}, {Darling}, {Davidson}, {Davies}, {Dawson}, {De Lee}, {Diamond-Stanic}, {Cano-D{\'\i}az}, {S{\'a}nchez},
  {Donor}, {Duckworth}, {Dwelly}, {Eisenstein}, {Elsworth}, {Emsellem}, {Eracleous}, {Escoffier}, {Fan}, {Farr}, {Feng}, {Fern{\'a}ndez-Trincado}, {Feuillet}, {Filipp}, {Fillingham}, {Frinchaboy}, {Fromenteau}, {Galbany}, {Garc{\'\i}a}, {Garc{\'\i}a-Hern{\'a}ndez}, {Ge}, {Geisler}, {Gelfand}, {G{\'e}ron}, {Gibson}, {Goddy}, {Godoy-Rivera}, {Grabowski}, {Green}, {Greener}, {Grier}, {Griffith}, {Guo}, {Guy}, {Hadjara}, {Harding}, {Hasselquist}, {Hayes}, {Hearty}, {Hern{\'a}ndez}, {Hill}, {Hogg}, {Holtzman}, {Horta}, {Hsieh}, {Hsu}, {Hsu}, {Huber}, {Huertas-Company}, {Hutchinson}, {Hwang}, {Ibarra-Medel}, {Chitham}, {Ilha}, {Imig}, {Jaekle}, {Jayasinghe}, {Ji}, {Johnson}, {Jones}, {J{\"o}nsson}, {Katkov}, {Khalatyan}, {Kinemuchi}, {Kisku}, {Knapen}, {Kneib}, {Kollmeier}, {Kong}, {Kounkel}, {Kreckel}, {Krishnarao}, {Lacerna}, {Lane}, {Langgin}, {Lavender}, {Law}, {Lazarz}, {Leung}, {Leung}, {Lewis}, {Li}, {Li}, {Lian}, {Liang}, {Lin}, {Lin}, {Lin}, {Lintott}, {Long}, {Longa-Pe{\~n}a}, {L{\'o}pez-Cob{\'a}}, {Lu},
  {Lundgren}, {Luo}, {Mackereth}, {de la Macorra}, {Mahadevan}, {Majewski}, {Manchado}, {Mandeville}, {Maraston}, {Margalef-Bentabol}, {Masseron}, {Masters}, {Mathur}, {McDermid}, {Mckay}, {Merloni}, {Merrifield}, {Meszaros}, {Miglio}, {Di Mille}, {Minniti}, {Minsley}, {Monachesi}, {Moon}, {Mosser}, {Mulchaey}, {Muna}, {Mu{\~n}oz}, {Myers}, {Myers}, {Nadathur}, {Nair}, {Nandra}, {Neumann}, {Newman}, {Nidever}, {Nikakhtar}, {Nitschelm}, {O'Connell}, {Garma-Oehmichen}, {Luan Souza de Oliveira}, {Olney}, {Oravetz}, {Ortigoza-Urdaneta}, {Osorio}, {Otter}, {Pace}, {Padilla}, {Pan}, {Pan}, {Parikh}, {Parker}, {Peirani}, {Pe{\~n}a Ram{\'\i}rez}, {Penny}, {Percival}, {Perez-Fournon}, {Pinsonneault}, {Poidevin}, {Poovelil}, {Price-Whelan}, {B{\'a}rbara de Andrade Queiroz}, {Raddick}, {Ray}, {Rembold}, {Riddle}, {Riffel}, {Riffel}, {Rix}, {Robin}, {Rodr{\'\i}guez-Puebla}, {Roman-Lopes}, {Rom{\'a}n-Z{\'u}{\~n}iga}, {Rose}, {Ross}, {Rossi}, {Rubin}, {Salvato}, {S{\'a}nchez}, {S{\'a}nchez-Gallego}, {Sanderson}, {Santana
  Rojas}, {Sarceno}, {Sarmiento}, {Sayres}, {Sazonova}, {Schaefer}, {Schiavon}, {Schlegel}, {Schneider}, {Schultheis}, {Schwope}, {Serenelli}, {Serna}, {Shao}, {Shapiro}, {Sharma}, {Shen}, {Shetrone}, {Shu}, {Simon}, {Skrutskie}, {Smethurst}, {Smith}, {Sobeck}, {Spoo}, {Sprague}, {Stark}, {Stassun}, {Steinmetz}, {Stello}, {Stone-Martinez}, {Storchi-Bergmann}, {Stringfellow}, {Stutz}, {Su}, {Taghizadeh-Popp}, {Talbot}, {Tayar}, {Telles}, {Teske}, {Thakar}, {Theissen}, {Tkachenko}, {Thomas}, {Tojeiro}, {Hernandez Toledo}, {Troup}, {Trump}, {Trussler}, {Turner}, {Tuttle}, {Unda-Sanzana}, {V{\'a}zquez-Mata}, {Valentini}, {Valenzuela}, {Vargas-Gonz{\'a}lez}, {Vargas-Maga{\~n}a}, {Alfaro}, {Villanova}, {Vincenzo}, {Wake}, {Warfield}, {Washington}, {Weaver}, {Weijmans}, {Weinberg}, {Weiss}, {Westfall}, {Wild}, {Wilde}, {Wilson}, {Wilson}, {Wilson}, {Wolf}, {Wood-Vasey}, {Yan}, {Zamora}, {Zasowski}, {Zhang}, {Zhao}, {Zheng}, {Zheng} and {Zhu}}]{2022ApJS..259...35A}
\bibinfo{author}{{Abdurro'uf}}, \bibinfo{author}{{Accetta}, K.}, \bibinfo{author}{{Aerts}, C.}, \bibinfo{author}{{Silva Aguirre}, V.}, \bibinfo{author}{{Ahumada}, R.}, \bibinfo{author}{{Ajgaonkar}, N.}, \bibinfo{author}{{Filiz Ak}, N.}, \bibinfo{author}{{Alam}, S.}, \bibinfo{author}{{Allende Prieto}, C.}, \bibinfo{author}{{Almeida}, A.}, \bibinfo{author}{{Anders}, F.}, \bibinfo{author}{{Anderson}, S.F.}, \bibinfo{author}{{Andrews}, B.H.}, \bibinfo{author}{{Anguiano}, B.}, \bibinfo{author}{{Aquino-Ort{\'\i}z}, E.}, \bibinfo{author}{{Arag{\'o}n-Salamanca}, A.}, \bibinfo{author}{{Argudo-Fern{\'a}ndez}, M.}, \bibinfo{author}{{Ata}, M.}, \bibinfo{author}{{Aubert}, M.}, \bibinfo{author}{{Avila-Reese}, V.}, \bibinfo{author}{{Badenes}, C.}, \bibinfo{author}{{Barb{\'a}}, R.H.}, \bibinfo{author}{{Barger}, K.}, \bibinfo{author}{{Barrera-Ballesteros}, J.K.}, \bibinfo{author}{{Beaton}, R.L.}, \bibinfo{author}{{Beers}, T.C.}, \bibinfo{author}{{Belfiore}, F.}, \bibinfo{author}{{Bender}, C.F.}, \bibinfo{author}{{Bernardi},
  M.}, \bibinfo{author}{{Bershady}, M.A.}, \bibinfo{author}{{Beutler}, F.}, \bibinfo{author}{{Bidin}, C.M.}, \bibinfo{author}{{Bird}, J.C.}, \bibinfo{author}{{Bizyaev}, D.}, \bibinfo{author}{{Blanc}, G.A.}, \bibinfo{author}{{Blanton}, M.R.}, \bibinfo{author}{{Boardman}, N.F.}, \bibinfo{author}{{Bolton}, A.S.}, \bibinfo{author}{{Boquien}, M.}, \bibinfo{author}{{Borissova}, J.}, \bibinfo{author}{{Bovy}, J.}, \bibinfo{author}{{Brandt}, W.N.}, \bibinfo{author}{{Brown}, J.}, \bibinfo{author}{{Brownstein}, J.R.}, \bibinfo{author}{{Brusa}, M.}, \bibinfo{author}{{Buchner}, J.}, \bibinfo{author}{{Bundy}, K.}, \bibinfo{author}{{Burchett}, J.N.}, \bibinfo{author}{{Bureau}, M.}, \bibinfo{author}{{Burgasser}, A.}, \bibinfo{author}{{Cabang}, T.K.}, \bibinfo{author}{{Campbell}, S.}, \bibinfo{author}{{Cappellari}, M.}, \bibinfo{author}{{Carlberg}, J.K.}, \bibinfo{author}{{Wanderley}, F.C.}, \bibinfo{author}{{Carrera}, R.}, \bibinfo{author}{{Cash}, J.}, \bibinfo{author}{{Chen}, Y.P.}, \bibinfo{author}{{Chen}, W.H.},
  \bibinfo{author}{{Cherinka}, B.}, \bibinfo{author}{{Chiappini}, C.}, \bibinfo{author}{{Choi}, P.D.}, \bibinfo{author}{{Chojnowski}, S.D.}, \bibinfo{author}{{Chung}, H.}, \bibinfo{author}{{Clerc}, N.}, \bibinfo{author}{{Cohen}, R.E.}, \bibinfo{author}{{Comerford}, J.M.}, \bibinfo{author}{{Comparat}, J.}, \bibinfo{author}{{da Costa}, L.}, \bibinfo{author}{{Covey}, K.}, \bibinfo{author}{{Crane}, J.D.}, \bibinfo{author}{{Cruz-Gonzalez}, I.}, \bibinfo{author}{{Culhane}, C.}, \bibinfo{author}{{Cunha}, K.}, \bibinfo{author}{{Dai}, Y.S.}, \bibinfo{author}{{Damke}, G.}, \bibinfo{author}{{Darling}, J.}, \bibinfo{author}{{Davidson}, James~W., J.}, \bibinfo{author}{{Davies}, R.}, \bibinfo{author}{{Dawson}, K.}, \bibinfo{author}{{De Lee}, N.}, \bibinfo{author}{{Diamond-Stanic}, A.M.}, \bibinfo{author}{{Cano-D{\'\i}az}, M.}, \bibinfo{author}{{S{\'a}nchez}, H.D.}, \bibinfo{author}{{Donor}, J.}, \bibinfo{author}{{Duckworth}, C.}, \bibinfo{author}{{Dwelly}, T.}, \bibinfo{author}{{Eisenstein}, D.J.},
  \bibinfo{author}{{Elsworth}, Y.P.}, \bibinfo{author}{{Emsellem}, E.}, \bibinfo{author}{{Eracleous}, M.}, \bibinfo{author}{{Escoffier}, S.}, \bibinfo{author}{{Fan}, X.}, \bibinfo{author}{{Farr}, E.}, \bibinfo{author}{{Feng}, S.}, \bibinfo{author}{{Fern{\'a}ndez-Trincado}, J.G.}, \bibinfo{author}{{Feuillet}, D.}, \bibinfo{author}{{Filipp}, A.}, \bibinfo{author}{{Fillingham}, S.P.}, \bibinfo{author}{{Frinchaboy}, P.M.}, \bibinfo{author}{{Fromenteau}, S.}, \bibinfo{author}{{Galbany}, L.}, \bibinfo{author}{{Garc{\'\i}a}, R.A.}, \bibinfo{author}{{Garc{\'\i}a-Hern{\'a}ndez}, D.A.}, \bibinfo{author}{{Ge}, J.}, \bibinfo{author}{{Geisler}, D.}, \bibinfo{author}{{Gelfand}, J.}, \bibinfo{author}{{G{\'e}ron}, T.}, \bibinfo{author}{{Gibson}, B.J.}, \bibinfo{author}{{Goddy}, J.}, \bibinfo{author}{{Godoy-Rivera}, D.}, \bibinfo{author}{{Grabowski}, K.}, \bibinfo{author}{{Green}, P.J.}, \bibinfo{author}{{Greener}, M.}, \bibinfo{author}{{Grier}, C.J.}, \bibinfo{author}{{Griffith}, E.}, \bibinfo{author}{{Guo}, H.},
  \bibinfo{author}{{Guy}, J.}, \bibinfo{author}{{Hadjara}, M.}, \bibinfo{author}{{Harding}, P.}, \bibinfo{author}{{Hasselquist}, S.}, \bibinfo{author}{{Hayes}, C.R.}, \bibinfo{author}{{Hearty}, F.}, \bibinfo{author}{{Hern{\'a}ndez}, J.}, \bibinfo{author}{{Hill}, L.}, \bibinfo{author}{{Hogg}, D.W.}, \bibinfo{author}{{Holtzman}, J.A.}, \bibinfo{author}{{Horta}, D.}, \bibinfo{author}{{Hsieh}, B.C.}, \bibinfo{author}{{Hsu}, C.H.}, \bibinfo{author}{{Hsu}, Y.H.}, \bibinfo{author}{{Huber}, D.}, \bibinfo{author}{{Huertas-Company}, M.}, \bibinfo{author}{{Hutchinson}, B.}, \bibinfo{author}{{Hwang}, H.S.}, \bibinfo{author}{{Ibarra-Medel}, H.J.}, \bibinfo{author}{{Chitham}, J.I.}, \bibinfo{author}{{Ilha}, G.S.}, \bibinfo{author}{{Imig}, J.}, \bibinfo{author}{{Jaekle}, W.}, \bibinfo{author}{{Jayasinghe}, T.}, \bibinfo{author}{{Ji}, X.}, \bibinfo{author}{{Johnson}, J.A.}, \bibinfo{author}{{Jones}, A.}, \bibinfo{author}{{J{\"o}nsson}, H.}, \bibinfo{author}{{Katkov}, I.}, \bibinfo{author}{{Khalatyan}, Arman, D.},
  \bibinfo{author}{{Kinemuchi}, K.}, \bibinfo{author}{{Kisku}, S.}, \bibinfo{author}{{Knapen}, J.H.}, \bibinfo{author}{{Kneib}, J.P.}, \bibinfo{author}{{Kollmeier}, J.A.}, \bibinfo{author}{{Kong}, M.}, \bibinfo{author}{{Kounkel}, M.}, \bibinfo{author}{{Kreckel}, K.}, \bibinfo{author}{{Krishnarao}, D.}, \bibinfo{author}{{Lacerna}, I.}, \bibinfo{author}{{Lane}, R.R.}, \bibinfo{author}{{Langgin}, R.}, \bibinfo{author}{{Lavender}, R.}, \bibinfo{author}{{Law}, D.R.}, \bibinfo{author}{{Lazarz}, D.}, \bibinfo{author}{{Leung}, H.W.}, \bibinfo{author}{{Leung}, H.H.}, \bibinfo{author}{{Lewis}, H.M.}, \bibinfo{author}{{Li}, C.}, \bibinfo{author}{{Li}, R.}, \bibinfo{author}{{Lian}, J.}, \bibinfo{author}{{Liang}, F.H.}, \bibinfo{author}{{Lin}, L.}, \bibinfo{author}{{Lin}, Y.T.}, \bibinfo{author}{{Lin}, S.}, \bibinfo{author}{{Lintott}, C.}, \bibinfo{author}{{Long}, D.}, \bibinfo{author}{{Longa-Pe{\~n}a}, P.}, \bibinfo{author}{{L{\'o}pez-Cob{\'a}}, C.}, \bibinfo{author}{{Lu}, S.}, \bibinfo{author}{{Lundgren}, B.F.},
  \bibinfo{author}{{Luo}, Y.}, \bibinfo{author}{{Mackereth}, J.T.}, \bibinfo{author}{{de la Macorra}, A.}, \bibinfo{author}{{Mahadevan}, S.}, \bibinfo{author}{{Majewski}, S.R.}, \bibinfo{author}{{Manchado}, A.}, \bibinfo{author}{{Mandeville}, T.}, \bibinfo{author}{{Maraston}, C.}, \bibinfo{author}{{Margalef-Bentabol}, B.}, \bibinfo{author}{{Masseron}, T.}, \bibinfo{author}{{Masters}, K.L.}, \bibinfo{author}{{Mathur}, S.}, \bibinfo{author}{{McDermid}, R.M.}, \bibinfo{author}{{Mckay}, M.}, \bibinfo{author}{{Merloni}, A.}, \bibinfo{author}{{Merrifield}, M.}, \bibinfo{author}{{Meszaros}, S.}, \bibinfo{author}{{Miglio}, A.}, \bibinfo{author}{{Di Mille}, F.}, \bibinfo{author}{{Minniti}, D.}, \bibinfo{author}{{Minsley}, R.}, \bibinfo{author}{{Monachesi}, A.}, \bibinfo{author}{{Moon}, J.}, \bibinfo{author}{{Mosser}, B.}, \bibinfo{author}{{Mulchaey}, J.}, \bibinfo{author}{{Muna}, D.}, \bibinfo{author}{{Mu{\~n}oz}, R.R.}, \bibinfo{author}{{Myers}, A.D.}, \bibinfo{author}{{Myers}, N.}, \bibinfo{author}{{Nadathur}, S.},
  \bibinfo{author}{{Nair}, P.}, \bibinfo{author}{{Nandra}, K.}, \bibinfo{author}{{Neumann}, J.}, \bibinfo{author}{{Newman}, J.A.}, \bibinfo{author}{{Nidever}, D.L.}, \bibinfo{author}{{Nikakhtar}, F.}, \bibinfo{author}{{Nitschelm}, C.}, \bibinfo{author}{{O'Connell}, J.E.}, \bibinfo{author}{{Garma-Oehmichen}, L.}, \bibinfo{author}{{Luan Souza de Oliveira}, G.}, \bibinfo{author}{{Olney}, R.}, \bibinfo{author}{{Oravetz}, D.}, \bibinfo{author}{{Ortigoza-Urdaneta}, M.}, \bibinfo{author}{{Osorio}, Y.}, \bibinfo{author}{{Otter}, J.}, \bibinfo{author}{{Pace}, Z.J.}, \bibinfo{author}{{Padilla}, N.}, \bibinfo{author}{{Pan}, K.}, \bibinfo{author}{{Pan}, H.A.}, \bibinfo{author}{{Parikh}, T.}, \bibinfo{author}{{Parker}, J.}, \bibinfo{author}{{Peirani}, S.}, \bibinfo{author}{{Pe{\~n}a Ram{\'\i}rez}, K.}, \bibinfo{author}{{Penny}, S.}, \bibinfo{author}{{Percival}, W.J.}, \bibinfo{author}{{Perez-Fournon}, I.}, \bibinfo{author}{{Pinsonneault}, M.}, \bibinfo{author}{{Poidevin}, F.}, \bibinfo{author}{{Poovelil}, V.J.},
  \bibinfo{author}{{Price-Whelan}, A.M.}, \bibinfo{author}{{B{\'a}rbara de Andrade Queiroz}, A.}, \bibinfo{author}{{Raddick}, M.J.}, \bibinfo{author}{{Ray}, A.}, \bibinfo{author}{{Rembold}, S.B.}, \bibinfo{author}{{Riddle}, N.}, \bibinfo{author}{{Riffel}, R.A.}, \bibinfo{author}{{Riffel}, R.}, \bibinfo{author}{{Rix}, H.W.}, \bibinfo{author}{{Robin}, A.C.}, \bibinfo{author}{{Rodr{\'\i}guez-Puebla}, A.}, \bibinfo{author}{{Roman-Lopes}, A.}, \bibinfo{author}{{Rom{\'a}n-Z{\'u}{\~n}iga}, C.}, \bibinfo{author}{{Rose}, B.}, \bibinfo{author}{{Ross}, A.J.}, \bibinfo{author}{{Rossi}, G.}, \bibinfo{author}{{Rubin}, K.H.R.}, \bibinfo{author}{{Salvato}, M.}, \bibinfo{author}{{S{\'a}nchez}, S.F.}, \bibinfo{author}{{S{\'a}nchez-Gallego}, J.R.}, \bibinfo{author}{{Sanderson}, R.}, \bibinfo{author}{{Santana Rojas}, F.A.}, \bibinfo{author}{{Sarceno}, E.}, \bibinfo{author}{{Sarmiento}, R.}, \bibinfo{author}{{Sayres}, C.}, \bibinfo{author}{{Sazonova}, E.}, \bibinfo{author}{{Schaefer}, A.L.}, \bibinfo{author}{{Schiavon}, R.},
  \bibinfo{author}{{Schlegel}, D.J.}, \bibinfo{author}{{Schneider}, D.P.}, \bibinfo{author}{{Schultheis}, M.}, \bibinfo{author}{{Schwope}, A.}, \bibinfo{author}{{Serenelli}, A.}, \bibinfo{author}{{Serna}, J.}, \bibinfo{author}{{Shao}, Z.}, \bibinfo{author}{{Shapiro}, G.}, \bibinfo{author}{{Sharma}, A.}, \bibinfo{author}{{Shen}, Y.}, \bibinfo{author}{{Shetrone}, M.}, \bibinfo{author}{{Shu}, Y.}, \bibinfo{author}{{Simon}, J.D.}, \bibinfo{author}{{Skrutskie}, M.F.}, \bibinfo{author}{{Smethurst}, R.}, \bibinfo{author}{{Smith}, V.}, \bibinfo{author}{{Sobeck}, J.}, \bibinfo{author}{{Spoo}, T.}, \bibinfo{author}{{Sprague}, D.}, \bibinfo{author}{{Stark}, D.V.}, \bibinfo{author}{{Stassun}, K.G.}, \bibinfo{author}{{Steinmetz}, M.}, \bibinfo{author}{{Stello}, D.}, \bibinfo{author}{{Stone-Martinez}, A.}, \bibinfo{author}{{Storchi-Bergmann}, T.}, \bibinfo{author}{{Stringfellow}, G.S.}, \bibinfo{author}{{Stutz}, A.}, \bibinfo{author}{{Su}, Y.C.}, \bibinfo{author}{{Taghizadeh-Popp}, M.}, \bibinfo{author}{{Talbot}, M.S.},
  \bibinfo{author}{{Tayar}, J.}, \bibinfo{author}{{Telles}, E.}, \bibinfo{author}{{Teske}, J.}, \bibinfo{author}{{Thakar}, A.}, \bibinfo{author}{{Theissen}, C.}, \bibinfo{author}{{Tkachenko}, A.}, \bibinfo{author}{{Thomas}, D.}, \bibinfo{author}{{Tojeiro}, R.}, \bibinfo{author}{{Hernandez Toledo}, H.}, \bibinfo{author}{{Troup}, N.W.}, \bibinfo{author}{{Trump}, J.R.}, \bibinfo{author}{{Trussler}, J.}, \bibinfo{author}{{Turner}, J.}, \bibinfo{author}{{Tuttle}, S.}, \bibinfo{author}{{Unda-Sanzana}, E.}, \bibinfo{author}{{V{\'a}zquez-Mata}, J.A.}, \bibinfo{author}{{Valentini}, M.}, \bibinfo{author}{{Valenzuela}, O.}, \bibinfo{author}{{Vargas-Gonz{\'a}lez}, J.}, \bibinfo{author}{{Vargas-Maga{\~n}a}, M.}, \bibinfo{author}{{Alfaro}, P.V.}, \bibinfo{author}{{Villanova}, S.}, \bibinfo{author}{{Vincenzo}, F.}, \bibinfo{author}{{Wake}, D.}, \bibinfo{author}{{Warfield}, J.T.}, \bibinfo{author}{{Washington}, J.D.}, \bibinfo{author}{{Weaver}, B.A.}, \bibinfo{author}{{Weijmans}, A.M.}, \bibinfo{author}{{Weinberg}, D.H.},
  \bibinfo{author}{{Weiss}, A.}, \bibinfo{author}{{Westfall}, K.B.}, \bibinfo{author}{{Wild}, V.}, \bibinfo{author}{{Wilde}, M.C.}, \bibinfo{author}{{Wilson}, J.C.}, \bibinfo{author}{{Wilson}, R.F.}, \bibinfo{author}{{Wilson}, M.}, \bibinfo{author}{{Wolf}, J.}, \bibinfo{author}{{Wood-Vasey}, W.M.}, \bibinfo{author}{{Yan}, R.}, \bibinfo{author}{{Zamora}, O.}, \bibinfo{author}{{Zasowski}, G.}, \bibinfo{author}{{Zhang}, K.}, \bibinfo{author}{{Zhao}, C.}, \bibinfo{author}{{Zheng}, Z.}, \bibinfo{author}{{Zheng}, Z.}, \bibinfo{author}{{Zhu}, K.}, \bibinfo{year}{2022}.
\newblock \bibinfo{title}{{The Seventeenth Data Release of the Sloan Digital Sky Surveys: Complete Release of MaNGA, MaStar, and APOGEE-2 Data}}.
\newblock \bibinfo{journal}{\apjs} \bibinfo{volume}{259}, \bibinfo{pages}{35}.
\newblock \DOIprefix\doi{10.3847/1538-4365/ac4414}, \href{http://arxiv.org/abs/2112.02026}{{\tt arXiv:2112.02026}}.
%Type = Article
\bibitem[{{Alvarez} et~al.(2022){Alvarez}, {Randall}, {Su}, {Sarkar}, {Walker}, {Lee}, {Sarazin} and {Blanton}}]{2022ApJ...938...51A}
\bibinfo{author}{{Alvarez}, G.E.}, \bibinfo{author}{{Randall}, S.W.}, \bibinfo{author}{{Su}, Y.}, \bibinfo{author}{{Sarkar}, A.}, \bibinfo{author}{{Walker}, S.}, \bibinfo{author}{{Lee}, N.P.}, \bibinfo{author}{{Sarazin}, C.L.}, \bibinfo{author}{{Blanton}, E.}, \bibinfo{year}{2022}.
\newblock \bibinfo{title}{{Suzaku Observations of the Cluster Outskirts and Intercluster Filament in the Triple Merger Cluster A98}}.
\newblock \bibinfo{journal}{\apj} \bibinfo{volume}{938}, \bibinfo{pages}{51}.
\newblock \DOIprefix\doi{10.3847/1538-4357/ac91d3}, \href{http://arxiv.org/abs/2206.08430}{{\tt arXiv:2206.08430}}.
%Type = Inproceedings
\bibitem[{{Arnaud}(1996)}]{1996ASPC..101...17A}
\bibinfo{author}{{Arnaud}, K.A.}, \bibinfo{year}{1996}.
\newblock \bibinfo{title}{{XSPEC: The First Ten Years}}, in: \bibinfo{editor}{{Jacoby}, G.H.}, \bibinfo{editor}{{Barnes}, J.} (Eds.), \bibinfo{booktitle}{Astronomical Data Analysis Software and Systems V}, p.~\bibinfo{pages}{17}.
%Type = Article
\bibitem[{{Ascasibar} and {Markevitch}(2006)}]{2006ApJ...650..102A}
\bibinfo{author}{{Ascasibar}, Y.}, \bibinfo{author}{{Markevitch}, M.}, \bibinfo{year}{2006}.
\newblock \bibinfo{title}{{The Origin of Cold Fronts in the Cores of Relaxed Galaxy Clusters}}.
\newblock \bibinfo{journal}{\apj} \bibinfo{volume}{650}, \bibinfo{pages}{102--127}.
\newblock \DOIprefix\doi{10.1086/506508}, \href{http://arxiv.org/abs/astro-ph/0603246}{{\tt arXiv:astro-ph/0603246}}.
%Type = Article
\bibitem[{{Blanton} et~al.(2011){Blanton}, {Randall}, {Clarke}, {Sarazin}, {McNamara}, {Douglass} and {McDonald}}]{2011ApJ...737...99B}
\bibinfo{author}{{Blanton}, E.L.}, \bibinfo{author}{{Randall}, S.W.}, \bibinfo{author}{{Clarke}, T.E.}, \bibinfo{author}{{Sarazin}, C.L.}, \bibinfo{author}{{McNamara}, B.R.}, \bibinfo{author}{{Douglass}, E.M.}, \bibinfo{author}{{McDonald}, M.}, \bibinfo{year}{2011}.
\newblock \bibinfo{title}{{A Very Deep Chandra Observation of A2052: Bubbles, Shocks, and Sloshing}}.
\newblock \bibinfo{journal}{\apj} \bibinfo{volume}{737}, \bibinfo{pages}{99}.
\newblock \DOIprefix\doi{10.1088/0004-637X/737/2/99}, \href{http://arxiv.org/abs/1105.4572}{{\tt arXiv:1105.4572}}.
%Type = Article
\bibitem[{{Boehringer} and {Hensler}(1989)}]{1989A&A...215..147B}
\bibinfo{author}{{Boehringer}, H.}, \bibinfo{author}{{Hensler}, G.}, \bibinfo{year}{1989}.
\newblock \bibinfo{title}{{Metallicity-dependence of radiative cooling in optically thin, hot plasmas}}.
\newblock \bibinfo{journal}{\aap} \bibinfo{volume}{215}, \bibinfo{pages}{147--149}.
%Type = Article
\bibitem[{{Bykov} et~al.(2015){Bykov}, {Churazov}, {Ferrari}, {Forman}, {Kaastra}, {Klein}, {Markevitch} and {de Plaa}}]{2015SSRv..188..141B}
\bibinfo{author}{{Bykov}, A.M.}, \bibinfo{author}{{Churazov}, E.M.}, \bibinfo{author}{{Ferrari}, C.}, \bibinfo{author}{{Forman}, W.R.}, \bibinfo{author}{{Kaastra}, J.S.}, \bibinfo{author}{{Klein}, U.}, \bibinfo{author}{{Markevitch}, M.}, \bibinfo{author}{{de Plaa}, J.}, \bibinfo{year}{2015}.
\newblock \bibinfo{title}{{Structures and Components in Galaxy Clusters: Observations and Models}}.
\newblock \bibinfo{journal}{\ssr} \bibinfo{volume}{188}, \bibinfo{pages}{141--185}.
\newblock \DOIprefix\doi{10.1007/s11214-014-0129-4}, \href{http://arxiv.org/abs/1512.01456}{{\tt arXiv:1512.01456}}.
%Type = Article
\bibitem[{{Calzadilla} et~al.(2019){Calzadilla}, {Russell}, {McDonald}, {Fabian}, {Baum}, {Combes}, {Donahue}, {Edge}, {McNamara}, {Nulsen}, {O'Dea}, {Oonk}, {Tremblay} and {Vantyghem}}]{2019ApJ...875...65C}
\bibinfo{author}{{Calzadilla}, M.S.}, \bibinfo{author}{{Russell}, H.R.}, \bibinfo{author}{{McDonald}, M.A.}, \bibinfo{author}{{Fabian}, A.C.}, \bibinfo{author}{{Baum}, S.A.}, \bibinfo{author}{{Combes}, F.}, \bibinfo{author}{{Donahue}, M.}, \bibinfo{author}{{Edge}, A.C.}, \bibinfo{author}{{McNamara}, B.R.}, \bibinfo{author}{{Nulsen}, P.E.J.}, \bibinfo{author}{{O'Dea}, C.P.}, \bibinfo{author}{{Oonk}, J.B.R.}, \bibinfo{author}{{Tremblay}, G.R.}, \bibinfo{author}{{Vantyghem}, A.N.}, \bibinfo{year}{2019}.
\newblock \bibinfo{title}{{Revealing a Highly Dynamic Cluster Core in Abell 1664 with Chandra}}.
\newblock \bibinfo{journal}{\apj} \bibinfo{volume}{875}, \bibinfo{pages}{65}.
\newblock \DOIprefix\doi{10.3847/1538-4357/ab09f6}, \href{http://arxiv.org/abs/1810.00881}{{\tt arXiv:1810.00881}}.
%Type = Article
\bibitem[{{Canning} et~al.(2011){Canning}, {Fabian}, {Johnstone}, {Sanders}, {Crawford}, {Ferland} and {Hatch}}]{2011MNRAS.417.3080C}
\bibinfo{author}{{Canning}, R.E.A.}, \bibinfo{author}{{Fabian}, A.C.}, \bibinfo{author}{{Johnstone}, R.M.}, \bibinfo{author}{{Sanders}, J.S.}, \bibinfo{author}{{Crawford}, C.S.}, \bibinfo{author}{{Ferland}, G.J.}, \bibinfo{author}{{Hatch}, N.A.}, \bibinfo{year}{2011}.
\newblock \bibinfo{title}{{A deep spectroscopic study of the filamentary nebulosity in NGC 4696, the brightest cluster galaxy in the Centaurus cluster}}.
\newblock \bibinfo{journal}{\mnras} \bibinfo{volume}{417}, \bibinfo{pages}{3080--3099}.
\newblock \DOIprefix\doi{10.1111/j.1365-2966.2011.19470.x}, \href{http://arxiv.org/abs/1107.4011}{{\tt arXiv:1107.4011}}.
%Type = Article
\bibitem[{{Cavaliere} and {Fusco-Femiano}(1976)}]{1976A&A....49..137C}
\bibinfo{author}{{Cavaliere}, A.}, \bibinfo{author}{{Fusco-Femiano}, R.}, \bibinfo{year}{1976}.
\newblock \bibinfo{title}{{X-rays from hot plasma in clusters of galaxies}}.
\newblock \bibinfo{journal}{\aap} \bibinfo{volume}{49}, \bibinfo{pages}{137--144}.
%Type = Article
\bibitem[{{Churazov} et~al.(2003){Churazov}, {Forman}, {Jones} and {B{\"o}hringer}}]{2003ApJ...590..225C}
\bibinfo{author}{{Churazov}, E.}, \bibinfo{author}{{Forman}, W.}, \bibinfo{author}{{Jones}, C.}, \bibinfo{author}{{B{\"o}hringer}, H.}, \bibinfo{year}{2003}.
\newblock \bibinfo{title}{{XMM-Newton Observations of the Perseus Cluster. I. The Temperature and Surface Brightness Structure}}.
\newblock \bibinfo{journal}{\apj} \bibinfo{volume}{590}, \bibinfo{pages}{225--237}.
\newblock \DOIprefix\doi{10.1086/374923}, \href{http://arxiv.org/abs/arXiv:astro-ph/0301482}{{\tt arXiv:arXiv:astro-ph/0301482}}.
%Type = Article
\bibitem[{{Douglass} et~al.(2018){Douglass}, {Blanton}, {Randall}, {Clarke}, {Edwards}, {Sabry} and {ZuHone}}]{2018ApJ...868..121D}
\bibinfo{author}{{Douglass}, E.M.}, \bibinfo{author}{{Blanton}, E.L.}, \bibinfo{author}{{Randall}, S.W.}, \bibinfo{author}{{Clarke}, T.E.}, \bibinfo{author}{{Edwards}, L.O.V.}, \bibinfo{author}{{Sabry}, Z.}, \bibinfo{author}{{ZuHone}, J.A.}, \bibinfo{year}{2018}.
\newblock \bibinfo{title}{{The Megaparsec-scale Gas-sloshing Spiral in the Remnant Cool Core Cluster Abell 1763}}.
\newblock \bibinfo{journal}{\apj} \bibinfo{volume}{868}, \bibinfo{pages}{121}.
\newblock \DOIprefix\doi{10.3847/1538-4357/aae9e7}, \href{http://arxiv.org/abs/1812.02645}{{\tt arXiv:1812.02645}}.
%Type = Article
\bibitem[{{Dressler}(1980)}]{1980ApJ...236..351D}
\bibinfo{author}{{Dressler}, A.}, \bibinfo{year}{1980}.
\newblock \bibinfo{title}{{Galaxy morphology in rich clusters: implications for the formation and evolution of galaxies.}}
\newblock \bibinfo{journal}{\apj} \bibinfo{volume}{236}, \bibinfo{pages}{351--365}.
\newblock \DOIprefix\doi{10.1086/157753}.
%Type = Article
\bibitem[{{Dupke} et~al.(2007){Dupke}, {White} and {Bregman}}]{2007ApJ...671..181D}
\bibinfo{author}{{Dupke}, R.}, \bibinfo{author}{{White}, Raymond~E., I.}, \bibinfo{author}{{Bregman}, J.N.}, \bibinfo{year}{2007}.
\newblock \bibinfo{title}{{Different Methods of Forming Cold Fronts in Nonmerging Clusters}}.
\newblock \bibinfo{journal}{\apj} \bibinfo{volume}{671}, \bibinfo{pages}{181--189}.
\newblock \DOIprefix\doi{10.1086/522194}, \href{http://arxiv.org/abs/0707.4001}{{\tt arXiv:0707.4001}}.
%Type = Article
\bibitem[{{Dupke} and {White}(2003)}]{2003ApJ...583L..13D}
\bibinfo{author}{{Dupke}, R.}, \bibinfo{author}{{White}, III, R.E.}, \bibinfo{year}{2003}.
\newblock \bibinfo{title}{{Chandra Analysis of A496: No Chemical Gradients across Cold Fronts}}.
\newblock \bibinfo{journal}{\apjl} \bibinfo{volume}{583}, \bibinfo{pages}{L13--L16}.
\newblock \DOIprefix\doi{10.1086/367824}, \href{http://arxiv.org/abs/arXiv:astro-ph/0212332}{{\tt arXiv:arXiv:astro-ph/0212332}}.
%Type = Article
\bibitem[{{Eckert} et~al.(2011){Eckert}, {Molendi} and {Paltani}}]{2011A&A...526A..79E}
\bibinfo{author}{{Eckert}, D.}, \bibinfo{author}{{Molendi}, S.}, \bibinfo{author}{{Paltani}, S.}, \bibinfo{year}{2011}.
\newblock \bibinfo{title}{{The cool-core bias in X-ray galaxy cluster samples. I. Method and application to HIFLUGCS}}.
\newblock \bibinfo{journal}{\aap} \bibinfo{volume}{526}, \bibinfo{pages}{A79}.
\newblock \DOIprefix\doi{10.1051/0004-6361/201015856}, \href{http://arxiv.org/abs/1011.3302}{{\tt arXiv:1011.3302}}.
%Type = Inproceedings
\bibitem[{{Gastaldello} et~al.(2009){Gastaldello}, {Buote}, {Brighenti}, {Mathews}, {Temi} and {Ettori}}]{2009AIPC.1201..237G}
\bibinfo{author}{{Gastaldello}, F.}, \bibinfo{author}{{Buote}, D.A.}, \bibinfo{author}{{Brighenti}, F.}, \bibinfo{author}{{Mathews}, W.G.}, \bibinfo{author}{{Temi}, P.}, \bibinfo{author}{{Ettori}, S.}, \bibinfo{year}{2009}.
\newblock \bibinfo{title}{{AGN Feedback in Galaxy Groups: The Two Interesting Cases of AWM 4 and NGC 5044}}, in: \bibinfo{editor}{{Heinz}, S.}, \bibinfo{editor}{{Wilcots}, E.} (Eds.), \bibinfo{booktitle}{The Monster's Fiery Breath: Feedback in Galaxies, Groups, and Clusters}, pp. \bibinfo{pages}{237--240}.
\newblock \DOIprefix\doi{10.1063/1.3293045}, \href{http://arxiv.org/abs/0909.0600}{{\tt arXiv:0909.0600}}.
%Type = Article
\bibitem[{Gopal-Krishna et~al.(2022)Gopal-Krishna, Paul, Salunkhe and Sonkamble}]{gopal-krishna_paul_salunkhe_sonkamble_2022}
\bibinfo{author}{Gopal-Krishna}, \bibinfo{author}{Paul, S.}, \bibinfo{author}{Salunkhe, S.}, \bibinfo{author}{Sonkamble, S.}, \bibinfo{year}{2022}.
\newblock \bibinfo{title}{The radio source in abell 980: A detached-double-double radio galaxy?}
\newblock \bibinfo{journal}{Publications of the Astronomical Society of Australia} \bibinfo{volume}{39}, \bibinfo{pages}{e049}.
\newblock \DOIprefix\doi{10.1017/pasa.2022.30}.
%Type = Article
\bibitem[{{Grevesse} and {Sauval}(1998)}]{1998SSRv...85..161G}
\bibinfo{author}{{Grevesse}, N.}, \bibinfo{author}{{Sauval}, A.J.}, \bibinfo{year}{1998}.
\newblock \bibinfo{title}{{Standard Solar Composition}}.
\newblock \bibinfo{journal}{\ssr} \bibinfo{volume}{85}, \bibinfo{pages}{161--174}.
\newblock \DOIprefix\doi{10.1023/A:1005161325181}.
%Type = Article
\bibitem[{{Hamer} et~al.(2016){Hamer}, {Edge}, {Swinbank}, {Wilman}, {Combes}, {Salom{\'e}}, {Fabian}, {Crawford}, {Russell}, {Hlavacek-Larrondo}, {McNamara} and {Bremer}}]{2016MNRAS.460.1758H}
\bibinfo{author}{{Hamer}, S.L.}, \bibinfo{author}{{Edge}, A.C.}, \bibinfo{author}{{Swinbank}, A.M.}, \bibinfo{author}{{Wilman}, R.J.}, \bibinfo{author}{{Combes}, F.}, \bibinfo{author}{{Salom{\'e}}, P.}, \bibinfo{author}{{Fabian}, A.C.}, \bibinfo{author}{{Crawford}, C.S.}, \bibinfo{author}{{Russell}, H.R.}, \bibinfo{author}{{Hlavacek-Larrondo}, J.}, \bibinfo{author}{{McNamara}, B.R.}, \bibinfo{author}{{Bremer}, M.N.}, \bibinfo{year}{2016}.
\newblock \bibinfo{title}{{Optical emission line nebulae in galaxy cluster cores 1: the morphological, kinematic and spectral properties of the sample}}.
\newblock \bibinfo{journal}{\mnras} \bibinfo{volume}{460}, \bibinfo{pages}{1758--1789}.
\newblock \DOIprefix\doi{10.1093/mnras/stw1054}, \href{http://arxiv.org/abs/1603.03047}{{\tt arXiv:1603.03047}}.
%Type = Article
\bibitem[{{Hamer} et~al.(2012){Hamer}, {Edge}, {Swinbank}, {Wilman}, {Russell}, {Fabian}, {Sanders} and {Salom{\'e}}}]{2012MNRAS.421.3409H}
\bibinfo{author}{{Hamer}, S.L.}, \bibinfo{author}{{Edge}, A.C.}, \bibinfo{author}{{Swinbank}, A.M.}, \bibinfo{author}{{Wilman}, R.J.}, \bibinfo{author}{{Russell}, H.R.}, \bibinfo{author}{{Fabian}, A.C.}, \bibinfo{author}{{Sanders}, J.S.}, \bibinfo{author}{{Salom{\'e}}, P.}, \bibinfo{year}{2012}.
\newblock \bibinfo{title}{{The relation between line emission and brightest cluster galaxies in three exceptional clusters: evidence for gas cooling from the intracluster medium}}.
\newblock \bibinfo{journal}{\mnras} \bibinfo{volume}{421}, \bibinfo{pages}{3409--3417}.
\newblock \DOIprefix\doi{10.1111/j.1365-2966.2012.20566.x}, \href{http://arxiv.org/abs/1112.4848}{{\tt arXiv:1112.4848}}.
%Type = Article
\bibitem[{{Ichinohe} et~al.(2019){Ichinohe}, {Simionescu}, {Werner}, {Fabian} and {Takahashi}}]{2019MNRAS.483.1744I}
\bibinfo{author}{{Ichinohe}, Y.}, \bibinfo{author}{{Simionescu}, A.}, \bibinfo{author}{{Werner}, N.}, \bibinfo{author}{{Fabian}, A.C.}, \bibinfo{author}{{Takahashi}, T.}, \bibinfo{year}{2019}.
\newblock \bibinfo{title}{{Substructures associated with the sloshing cold front in the Perseus cluster}}.
\newblock \bibinfo{journal}{\mnras} \bibinfo{volume}{483}, \bibinfo{pages}{1744--1753}.
\newblock \DOIprefix\doi{10.1093/mnras/sty3257}, \href{http://arxiv.org/abs/1810.07380}{{\tt arXiv:1810.07380}}.
%Type = Article
\bibitem[{{Jennings} and {Dav{\'e}}(2023)}]{2023MNRAS.526.1367J}
\bibinfo{author}{{Jennings}, F.}, \bibinfo{author}{{Dav{\'e}}, R.}, \bibinfo{year}{2023}.
\newblock \bibinfo{title}{{Halo scaling relations and hydrostatic mass bias in the SIMBA simulation from realistic mock X-ray catalogues}}.
\newblock \bibinfo{journal}{\mnras} \bibinfo{volume}{526}, \bibinfo{pages}{1367--1387}.
\newblock \DOIprefix\doi{10.1093/mnras/stad2666}, \href{http://arxiv.org/abs/2306.01397}{{\tt arXiv:2306.01397}}.
%Type = Phdthesis
\bibitem[{{Johnson}(2011)}]{2011PhDT........14J}
\bibinfo{author}{{Johnson}, R.}, \bibinfo{year}{2011}.
\newblock \bibinfo{title}{{The Nature of Core Gas Sloshing in Galaxy Clusters}}.
\newblock Ph.D. thesis. Dartmouth College, New Hampshire.
%Type = Article
\bibitem[{{Johnson} et~al.(2010){Johnson}, {Markevitch}, {Wegner}, {Jones} and {Forman}}]{2010ApJ...710.1776J}
\bibinfo{author}{{Johnson}, R.E.}, \bibinfo{author}{{Markevitch}, M.}, \bibinfo{author}{{Wegner}, G.A.}, \bibinfo{author}{{Jones}, C.}, \bibinfo{author}{{Forman}, W.R.}, \bibinfo{year}{2010}.
\newblock \bibinfo{title}{{Core Gas Sloshing in Abell 1644}}.
\newblock \bibinfo{journal}{\apj} \bibinfo{volume}{710}, \bibinfo{pages}{1776--1785}.
\newblock \DOIprefix\doi{10.1088/0004-637X/710/2/1776}, \href{http://arxiv.org/abs/1001.2441}{{\tt arXiv:1001.2441}}.
%Type = Article
\bibitem[{{Johnson} et~al.(2012){Johnson}, {Zuhone}, {Jones}, {Forman} and {Markevitch}}]{2012ApJ...751...95J}
\bibinfo{author}{{Johnson}, R.E.}, \bibinfo{author}{{Zuhone}, J.}, \bibinfo{author}{{Jones}, C.}, \bibinfo{author}{{Forman}, W.R.}, \bibinfo{author}{{Markevitch}, M.}, \bibinfo{year}{2012}.
\newblock \bibinfo{title}{{Sloshing Gas in the Core of the Most Luminous Galaxy Cluster RXJ1347.5-1145}}.
\newblock \bibinfo{journal}{\apj} \bibinfo{volume}{751}, \bibinfo{pages}{95}.
\newblock \DOIprefix\doi{10.1088/0004-637X/751/2/95}, \href{http://arxiv.org/abs/1106.3489}{{\tt arXiv:1106.3489}}.
%Type = Article
\bibitem[{{Kadam} et~al.(2019){Kadam}, {Sonkamble}, {Pawar} and {Patil}}]{2019MNRAS.484.4113K}
\bibinfo{author}{{Kadam}, S.K.}, \bibinfo{author}{{Sonkamble}, S.S.}, \bibinfo{author}{{Pawar}, P.K.}, \bibinfo{author}{{Patil}, M.K.}, \bibinfo{year}{2019}.
\newblock \bibinfo{title}{{Merging cold front and AGN feedback in the peculiar galaxy cluster Abell 2626}}.
\newblock \bibinfo{journal}{\mnras} \bibinfo{volume}{484}, \bibinfo{pages}{4113--4126}.
\newblock \DOIprefix\doi{10.1093/mnras/stz144}, \href{http://arxiv.org/abs/1901.03550}{{\tt arXiv:1901.03550}}.
%Type = Article
\bibitem[{{Kalberla} et~al.(2005){Kalberla}, {Burton}, {Hartmann}, {Arnal}, {Bajaja}, {Morras} and {Poeppel}}]{2005yCat.8076....0K}
\bibinfo{author}{{Kalberla}, P.M.W.}, \bibinfo{author}{{Burton}, W.B.}, \bibinfo{author}{{Hartmann}, D.}, \bibinfo{author}{{Arnal}, E.M.}, \bibinfo{author}{{Bajaja}, E.}, \bibinfo{author}{{Morras}, R.}, \bibinfo{author}{{Poeppel}, W.G.L.}, \bibinfo{year}{2005}.
\newblock \bibinfo{title}{{Leiden/Argentine/Bonn (LAB) Survey of Galactic HI (Kalberla+ 2005)}}.
\newblock \bibinfo{journal}{VizieR Online Data Catalog} \bibinfo{volume}{8076}, \bibinfo{pages}{0--+}.
%Type = Article
\bibitem[{{Lovisari} and {Reiprich}(2019)}]{2019MNRAS.483..540L}
\bibinfo{author}{{Lovisari}, L.}, \bibinfo{author}{{Reiprich}, T.H.}, \bibinfo{year}{2019}.
\newblock \bibinfo{title}{{The non-uniformity of galaxy cluster metallicity profiles}}.
\newblock \bibinfo{journal}{\mnras} \bibinfo{volume}{483}, \bibinfo{pages}{540--557}.
\newblock \DOIprefix\doi{10.1093/mnras/sty3130}, \href{http://arxiv.org/abs/1811.05987}{{\tt arXiv:1811.05987}}.
%Type = Article
\bibitem[{{Machacek} et~al.(2011){Machacek}, {Jerius}, {Kraft}, {Forman}, {Jones}, {Randall}, {Giacintucci} and {Sun}}]{2011ApJ...743...15M}
\bibinfo{author}{{Machacek}, M.E.}, \bibinfo{author}{{Jerius}, D.}, \bibinfo{author}{{Kraft}, R.}, \bibinfo{author}{{Forman}, W.R.}, \bibinfo{author}{{Jones}, C.}, \bibinfo{author}{{Randall}, S.}, \bibinfo{author}{{Giacintucci}, S.}, \bibinfo{author}{{Sun}, M.}, \bibinfo{year}{2011}.
\newblock \bibinfo{title}{{Deep Chandra Observations of Edges and Bubbles in the NGC 5846 Galaxy Group}}.
\newblock \bibinfo{journal}{\apj} \bibinfo{volume}{743}, \bibinfo{pages}{15}.
\newblock \DOIprefix\doi{10.1088/0004-637X/743/1/15}, \href{http://arxiv.org/abs/1108.5229}{{\tt arXiv:1108.5229}}.
%Type = Article
\bibitem[{{Mann} and {Ebeling}(2012)}]{2012MNRAS.420.2120M}
\bibinfo{author}{{Mann}, A.W.}, \bibinfo{author}{{Ebeling}, H.}, \bibinfo{year}{2012}.
\newblock \bibinfo{title}{{X-ray-optical classification of cluster mergers and the evolution of the cluster merger fraction}}.
\newblock \bibinfo{journal}{\mnras} \bibinfo{volume}{420}, \bibinfo{pages}{2120--2138}.
\newblock \DOIprefix\doi{10.1111/j.1365-2966.2011.20170.x}, \href{http://arxiv.org/abs/1111.2396}{{\tt arXiv:1111.2396}}.
%Type = Article
\bibitem[{{Markevitch} et~al.(2000){Markevitch}, {Ponman}, {Nulsen}, {Bautz}, {Burke}, {David}, {Davis}, {Donnelly}, {Forman}, {Jones}, {Kaastra}, {Kellogg}, {Kim}, {Kolodziejczak}, {Mazzotta}, {Pagliaro}, {Patel}, {Van Speybroeck}, {Vikhlinin}, {Vrtilek}, {Wise} and {Zhao}}]{2000ApJ...541..542M}
\bibinfo{author}{{Markevitch}, M.}, \bibinfo{author}{{Ponman}, T.J.}, \bibinfo{author}{{Nulsen}, P.E.J.}, \bibinfo{author}{{Bautz}, M.W.}, \bibinfo{author}{{Burke}, D.J.}, \bibinfo{author}{{David}, L.P.}, \bibinfo{author}{{Davis}, D.}, \bibinfo{author}{{Donnelly}, R.H.}, \bibinfo{author}{{Forman}, W.R.}, \bibinfo{author}{{Jones}, C.}, \bibinfo{author}{{Kaastra}, J.}, \bibinfo{author}{{Kellogg}, E.}, \bibinfo{author}{{Kim}, D.W.}, \bibinfo{author}{{Kolodziejczak}, J.}, \bibinfo{author}{{Mazzotta}, P.}, \bibinfo{author}{{Pagliaro}, A.}, \bibinfo{author}{{Patel}, S.}, \bibinfo{author}{{Van Speybroeck}, L.}, \bibinfo{author}{{Vikhlinin}, A.}, \bibinfo{author}{{Vrtilek}, J.}, \bibinfo{author}{{Wise}, M.}, \bibinfo{author}{{Zhao}, P.}, \bibinfo{year}{2000}.
\newblock \bibinfo{title}{{Chandra Observation of Abell 2142: Survival of Dense Subcluster Cores in a Merger}}.
\newblock \bibinfo{journal}{\apj} \bibinfo{volume}{541}, \bibinfo{pages}{542--549}.
\newblock \DOIprefix\doi{10.1086/309470}, \href{http://arxiv.org/abs/astro-ph/0001269}{{\tt arXiv:astro-ph/0001269}}.
%Type = Article
\bibitem[{{Markevitch} and {Vikhlinin}(2007)}]{2007PhR...443....1M}
\bibinfo{author}{{Markevitch}, M.}, \bibinfo{author}{{Vikhlinin}, A.}, \bibinfo{year}{2007}.
\newblock \bibinfo{title}{{Shocks and cold fronts in galaxy clusters}}.
\newblock \bibinfo{journal}{\physrep} \bibinfo{volume}{443}, \bibinfo{pages}{1--53}.
\newblock \DOIprefix\doi{10.1016/j.physrep.2007.01.001}, \href{http://arxiv.org/abs/astro-ph/0701821}{{\tt arXiv:astro-ph/0701821}}.
%Type = Article
\bibitem[{{Markevitch} et~al.(2001){Markevitch}, {Vikhlinin} and {Mazzotta}}]{2001ApJ...562L.153M}
\bibinfo{author}{{Markevitch}, M.}, \bibinfo{author}{{Vikhlinin}, A.}, \bibinfo{author}{{Mazzotta}, P.}, \bibinfo{year}{2001}.
\newblock \bibinfo{title}{{Nonhydrostatic Gas in the Core of the Relaxed Galaxy Cluster A1795}}.
\newblock \bibinfo{journal}{\apjl} \bibinfo{volume}{562}, \bibinfo{pages}{L153--L156}.
\newblock \DOIprefix\doi{10.1086/337973}, \href{http://arxiv.org/abs/arXiv:astro-ph/0108520}{{\tt arXiv:arXiv:astro-ph/0108520}}.
%Type = Article
\bibitem[{{Mazzotta} et~al.(2001){Mazzotta}, {Markevitch}, {Vikhlinin}, {Forman}, {David} and {van Speybroeck}}]{2001ApJ...555..205M}
\bibinfo{author}{{Mazzotta}, P.}, \bibinfo{author}{{Markevitch}, M.}, \bibinfo{author}{{Vikhlinin}, A.}, \bibinfo{author}{{Forman}, W.R.}, \bibinfo{author}{{David}, L.P.}, \bibinfo{author}{{van Speybroeck}, L.}, \bibinfo{year}{2001}.
\newblock \bibinfo{title}{{Chandra Observation of RX J1720.1+2638: a Nearly Relaxed Cluster with a Fast-moving Core?}}
\newblock \bibinfo{journal}{\apj} \bibinfo{volume}{555}, \bibinfo{pages}{205--214}.
\newblock \DOIprefix\doi{10.1086/321484}, \href{http://arxiv.org/abs/arXiv:astro-ph/0102291}{{\tt arXiv:arXiv:astro-ph/0102291}}.
%Type = Article
\bibitem[{{Owen} and {Ledlow}(1997)}]{1997ApJS..108...41O}
\bibinfo{author}{{Owen}, F.N.}, \bibinfo{author}{{Ledlow}, M.J.}, \bibinfo{year}{1997}.
\newblock \bibinfo{title}{{A 20 Centimeter VLA Survey of Abell Clusters of Galaxies. VII. Detailed Radio Images}}.
\newblock \bibinfo{journal}{\apjs} \bibinfo{volume}{108}, \bibinfo{pages}{41--98}.
\newblock \DOIprefix\doi{10.1086/312954}.
%Type = Article
\bibitem[{{Owers} et~al.(2009){Owers}, {Nulsen}, {Couch} and {Markevitch}}]{2009ApJ...704.1349O}
\bibinfo{author}{{Owers}, M.S.}, \bibinfo{author}{{Nulsen}, P.E.J.}, \bibinfo{author}{{Couch}, W.J.}, \bibinfo{author}{{Markevitch}, M.}, \bibinfo{year}{2009}.
\newblock \bibinfo{title}{{A High Fidelity Sample of Cold Front Clusters from the Chandra Archive}}.
\newblock \bibinfo{journal}{\apj} \bibinfo{volume}{704}, \bibinfo{pages}{1349--1370}.
\newblock \DOIprefix\doi{10.1088/0004-637X/704/2/1349}, \href{http://arxiv.org/abs/0909.2645}{{\tt arXiv:0909.2645}}.
%Type = Article
\bibitem[{{Pandge} et~al.(2017){Pandge}, {Bagchi}, {Sonkamble}, {Parekh}, {Patil}, {Dabhade}, {Navale}, {Raychaudhury} and {Jacob}}]{2017MNRAS.472.2042P}
\bibinfo{author}{{Pandge}, M.B.}, \bibinfo{author}{{Bagchi}, J.}, \bibinfo{author}{{Sonkamble}, S.S.}, \bibinfo{author}{{Parekh}, V.}, \bibinfo{author}{{Patil}, M.K.}, \bibinfo{author}{{Dabhade}, P.}, \bibinfo{author}{{Navale}, N.R.}, \bibinfo{author}{{Raychaudhury}, S.}, \bibinfo{author}{{Jacob}, J.}, \bibinfo{year}{2017}.
\newblock \bibinfo{title}{{MACS J0553.4-3342: a young merging galaxy cluster caught through the eyes of Chandra and HST}}.
\newblock \bibinfo{journal}{\mnras} \bibinfo{volume}{472}, \bibinfo{pages}{2042--2053}.
\newblock \DOIprefix\doi{10.1093/mnras/stx2028}, \href{http://arxiv.org/abs/1701.00197}{{\tt arXiv:1701.00197}}.
%Type = Article
\bibitem[{{Pandge} et~al.(2012){Pandge}, {Vagshette}, {David} and {Patil}}]{2012MNRAS.421..808P}
\bibinfo{author}{{Pandge}, M.B.}, \bibinfo{author}{{Vagshette}, N.D.}, \bibinfo{author}{{David}, L.P.}, \bibinfo{author}{{Patil}, M.K.}, \bibinfo{year}{2012}.
\newblock \bibinfo{title}{{Systematic study of X-ray cavities in the brightest galaxy in the Draco constellation NGC 6338}}.
\newblock \bibinfo{journal}{\mnras} \bibinfo{volume}{421}, \bibinfo{pages}{808--817}.
\newblock \DOIprefix\doi{10.1111/j.1365-2966.2011.20358.x}, \href{http://arxiv.org/abs/1202.1364}{{\tt arXiv:1202.1364}}.
%Type = Article
\bibitem[{{Pandge} et~al.(2013){Pandge}, {Vagshette}, {Sonkamble} and {Patil}}]{2013Ap&SS.345..183P}
\bibinfo{author}{{Pandge}, M.B.}, \bibinfo{author}{{Vagshette}, N.D.}, \bibinfo{author}{{Sonkamble}, S.S.}, \bibinfo{author}{{Patil}, M.K.}, \bibinfo{year}{2013}.
\newblock \bibinfo{title}{{Investigation of X-ray cavities in the cooling flow system Abell 1991}}.
\newblock \bibinfo{journal}{\apss} \bibinfo{volume}{345}, \bibinfo{pages}{183--193}.
\newblock \DOIprefix\doi{10.1007/s10509-013-1366-9}, \href{http://arxiv.org/abs/1301.2928}{{\tt arXiv:1301.2928}}.
%Type = Article
\bibitem[{{Pasini} et~al.(2021a){Pasini}, {Finoguenov}, {Br{\"u}ggen}, {Gaspari}, {de Gasperin} and {Gozaliasl}}]{2021MNRAS.505.2628P}
\bibinfo{author}{{Pasini}, T.}, \bibinfo{author}{{Finoguenov}, A.}, \bibinfo{author}{{Br{\"u}ggen}, M.}, \bibinfo{author}{{Gaspari}, M.}, \bibinfo{author}{{de Gasperin}, F.}, \bibinfo{author}{{Gozaliasl}, G.}, \bibinfo{year}{2021}a.
\newblock \bibinfo{title}{{Radio galaxies in galaxy groups: kinematics, scaling relations, and AGN feedback}}.
\newblock \bibinfo{journal}{\mnras} \bibinfo{volume}{505}, \bibinfo{pages}{2628--2637}.
\newblock \DOIprefix\doi{10.1093/mnras/stab1451}, \href{http://arxiv.org/abs/2105.08727}{{\tt arXiv:2105.08727}}.
%Type = Article
\bibitem[{{Pasini} et~al.(2021b){Pasini}, {Gitti}, {Brighenti}, {O'Sullivan}, {Gastaldello}, {Temi} and {Hamer}}]{2021ApJ...911...66P}
\bibinfo{author}{{Pasini}, T.}, \bibinfo{author}{{Gitti}, M.}, \bibinfo{author}{{Brighenti}, F.}, \bibinfo{author}{{O'Sullivan}, E.}, \bibinfo{author}{{Gastaldello}, F.}, \bibinfo{author}{{Temi}, P.}, \bibinfo{author}{{Hamer}, S.L.}, \bibinfo{year}{2021}b.
\newblock \bibinfo{title}{{A First Chandra View of the Cool Core Cluster A1668: Offset Cooling and AGN Feedback Cycle}}.
\newblock \bibinfo{journal}{\apj} \bibinfo{volume}{911}, \bibinfo{pages}{66}.
\newblock \DOIprefix\doi{10.3847/1538-4357/abe85f}, \href{http://arxiv.org/abs/2102.11299}{{\tt arXiv:2102.11299}}.
%Type = Article
\bibitem[{{Ponman} et~al.(2003){Ponman}, {Sanderson} and {Finoguenov}}]{2003MNRAS.343..331P}
\bibinfo{author}{{Ponman}, T.J.}, \bibinfo{author}{{Sanderson}, A.J.R.}, \bibinfo{author}{{Finoguenov}, A.}, \bibinfo{year}{2003}.
\newblock \bibinfo{title}{{The Birmingham-CfA cluster scaling project - III. Entropy and similarity in galaxy systems}}.
\newblock \bibinfo{journal}{\mnras} \bibinfo{volume}{343}, \bibinfo{pages}{331--342}.
\newblock \DOIprefix\doi{10.1046/j.1365-8711.2003.06677.x}, \href{http://arxiv.org/abs/astro-ph/0304048}{{\tt arXiv:astro-ph/0304048}}.
%Type = Article
\bibitem[{{Rahaman} et~al.(2022){Rahaman}, {Raja} and {Datta}}]{2022MNRAS.509.5821R}
\bibinfo{author}{{Rahaman}, M.}, \bibinfo{author}{{Raja}, R.}, \bibinfo{author}{{Datta}, A.}, \bibinfo{year}{2022}.
\newblock \bibinfo{title}{{On the detection of multiple shock fronts in A1914 using deep Chandra X-ray observations}}.
\newblock \bibinfo{journal}{\mnras} \bibinfo{volume}{509}, \bibinfo{pages}{5821--5835}.
\newblock \DOIprefix\doi{10.1093/mnras/stab3115}, \href{http://arxiv.org/abs/2110.12297}{{\tt arXiv:2110.12297}}.
%Type = Article
\bibitem[{{Randall} et~al.(2009a){Randall}, {Jones}, {Kraft}, {Forman} and {O'Sullivan}}]{2009ApJ...696.1431R}
\bibinfo{author}{{Randall}, S.W.}, \bibinfo{author}{{Jones}, C.}, \bibinfo{author}{{Kraft}, R.}, \bibinfo{author}{{Forman}, W.R.}, \bibinfo{author}{{O'Sullivan}, E.}, \bibinfo{year}{2009}a.
\newblock \bibinfo{title}{{Merging Cold Fronts in the Galaxy Pair NGC 7619 and NGC 7626}}.
\newblock \bibinfo{journal}{\apj} \bibinfo{volume}{696}, \bibinfo{pages}{1431--1440}.
\newblock \DOIprefix\doi{10.1088/0004-637X/696/2/1431}, \href{http://arxiv.org/abs/0811.1217}{{\tt arXiv:0811.1217}}.
%Type = Article
\bibitem[{{Randall} et~al.(2009b){Randall}, {Jones}, {Markevitch}, {Blanton}, {Nulsen} and {Forman}}]{2009ApJ...700.1404R}
\bibinfo{author}{{Randall}, S.W.}, \bibinfo{author}{{Jones}, C.}, \bibinfo{author}{{Markevitch}, M.}, \bibinfo{author}{{Blanton}, E.L.}, \bibinfo{author}{{Nulsen}, P.E.J.}, \bibinfo{author}{{Forman}, W.R.}, \bibinfo{year}{2009}b.
\newblock \bibinfo{title}{{Gas Sloshing and Bubbles in the Galaxy Group NGC 5098}}.
\newblock \bibinfo{journal}{\apj} \bibinfo{volume}{700}, \bibinfo{pages}{1404--1414}.
\newblock \DOIprefix\doi{10.1088/0004-637X/700/2/1404}, \href{http://arxiv.org/abs/0904.0610}{{\tt arXiv:0904.0610}}.
%Type = Article
\bibitem[{{Roediger} et~al.(2011){Roediger}, {Br{\"u}ggen}, {Simionescu}, {B{\"o}hringer}, {Churazov} and {Forman}}]{2011MNRAS.413.2057R}
\bibinfo{author}{{Roediger}, E.}, \bibinfo{author}{{Br{\"u}ggen}, M.}, \bibinfo{author}{{Simionescu}, A.}, \bibinfo{author}{{B{\"o}hringer}, H.}, \bibinfo{author}{{Churazov}, E.}, \bibinfo{author}{{Forman}, W.R.}, \bibinfo{year}{2011}.
\newblock \bibinfo{title}{{Gas sloshing, cold front formation and metal redistribution: the Virgo cluster as a quantitative test case}}.
\newblock \bibinfo{journal}{\mnras} \bibinfo{volume}{413}, \bibinfo{pages}{2057--2077}.
\newblock \DOIprefix\doi{10.1111/j.1365-2966.2011.18279.x}, \href{http://arxiv.org/abs/1007.4209}{{\tt arXiv:1007.4209}}.
%Type = Article
\bibitem[{{Rossetti} et~al.(2016){Rossetti}, {Gastaldello}, {Ferioli}, {Bersanelli}, {De Grandi}, {Eckert}, {Ghizzardi}, {Maino} and {Molendi}}]{2016MNRAS.457.4515R}
\bibinfo{author}{{Rossetti}, M.}, \bibinfo{author}{{Gastaldello}, F.}, \bibinfo{author}{{Ferioli}, G.}, \bibinfo{author}{{Bersanelli}, M.}, \bibinfo{author}{{De Grandi}, S.}, \bibinfo{author}{{Eckert}, D.}, \bibinfo{author}{{Ghizzardi}, S.}, \bibinfo{author}{{Maino}, D.}, \bibinfo{author}{{Molendi}, S.}, \bibinfo{year}{2016}.
\newblock \bibinfo{title}{{Measuring the dynamical state of Planck SZ-selected clusters: X-ray peak - BCG offset}}.
\newblock \bibinfo{journal}{\mnras} \bibinfo{volume}{457}, \bibinfo{pages}{4515--4524}.
\newblock \DOIprefix\doi{10.1093/mnras/stw265}, \href{http://arxiv.org/abs/1512.00410}{{\tt arXiv:1512.00410}}.
%Type = Article
\bibitem[{{Salunkhe} et~al.(2022){Salunkhe}, {Paul}, {Krishna}, {Sonkamble} and {Bhagat}}]{2022A&A...664A.186S}
\bibinfo{author}{{Salunkhe}, S.}, \bibinfo{author}{{Paul}, S.}, \bibinfo{author}{{Krishna}, G.}, \bibinfo{author}{{Sonkamble}, S.}, \bibinfo{author}{{Bhagat}, S.}, \bibinfo{year}{2022}.
\newblock \bibinfo{title}{{Deciphering the ultra-steep-spectrum diffuse radio sources discovered in the cool-core cluster Abell 980}}.
\newblock \bibinfo{journal}{\aap} \bibinfo{volume}{664}, \bibinfo{pages}{A186}.
\newblock \DOIprefix\doi{10.1051/0004-6361/202243438}, \href{http://arxiv.org/abs/2206.13550}{{\tt arXiv:2206.13550}}.
%Type = Article
\bibitem[{{Sanders}(2006)}]{2006MNRAS.371..829S}
\bibinfo{author}{{Sanders}, J.S.}, \bibinfo{year}{2006}.
\newblock \bibinfo{title}{{Contour binning: a new technique for spatially resolved X-ray spectroscopy applied to Cassiopeia A}}.
\newblock \bibinfo{journal}{\mnras} \bibinfo{volume}{371}, \bibinfo{pages}{829--842}.
\newblock \DOIprefix\doi{10.1111/j.1365-2966.2006.10716.x}, \href{http://arxiv.org/abs/arXiv:astro-ph/0606528}{{\tt arXiv:arXiv:astro-ph/0606528}}.
%Type = Article
\bibitem[{{Sanders} et~al.(2014){Sanders}, {Fabian}, {Hlavacek-Larrondo}, {Russell}, {Taylor}, {Hofmann}, {Tremblay} and {Walker}}]{2014MNRAS.444.1497S}
\bibinfo{author}{{Sanders}, J.S.}, \bibinfo{author}{{Fabian}, A.C.}, \bibinfo{author}{{Hlavacek-Larrondo}, J.}, \bibinfo{author}{{Russell}, H.R.}, \bibinfo{author}{{Taylor}, G.B.}, \bibinfo{author}{{Hofmann}, F.}, \bibinfo{author}{{Tremblay}, G.}, \bibinfo{author}{{Walker}, S.A.}, \bibinfo{year}{2014}.
\newblock \bibinfo{title}{{Feedback, scatter and structure in the core of the PKS 0745-191 galaxy cluster}}.
\newblock \bibinfo{journal}{\mnras} \bibinfo{volume}{444}, \bibinfo{pages}{1497--1517}.
\newblock \DOIprefix\doi{10.1093/mnras/stu1543}, \href{http://arxiv.org/abs/1407.8008}{{\tt arXiv:1407.8008}}.
%Type = Article
\bibitem[{{Sanders} et~al.(2016){Sanders}, {Fabian}, {Taylor}, {Russell}, {Blundell}, {Canning}, {Hlavacek-Larrondo}, {Walker} and {Grimes}}]{2016MNRAS.457...82S}
\bibinfo{author}{{Sanders}, J.S.}, \bibinfo{author}{{Fabian}, A.C.}, \bibinfo{author}{{Taylor}, G.B.}, \bibinfo{author}{{Russell}, H.R.}, \bibinfo{author}{{Blundell}, K.M.}, \bibinfo{author}{{Canning}, R.E.A.}, \bibinfo{author}{{Hlavacek-Larrondo}, J.}, \bibinfo{author}{{Walker}, S.A.}, \bibinfo{author}{{Grimes}, C.K.}, \bibinfo{year}{2016}.
\newblock \bibinfo{title}{{A very deep Chandra view of metals, sloshing and feedback in the Centaurus cluster of galaxies}}.
\newblock \bibinfo{journal}{\mnras} \bibinfo{volume}{457}, \bibinfo{pages}{82--109}.
\newblock \DOIprefix\doi{10.1093/mnras/stv2972}, \href{http://arxiv.org/abs/1601.01489}{{\tt arXiv:1601.01489}}.
%Type = Article
\bibitem[{{Sonkamble} et~al.(2015){Sonkamble}, {Vagshette}, {Pawar} and {Patil}}]{2015Ap&SS.359...61S}
\bibinfo{author}{{Sonkamble}, S.S.}, \bibinfo{author}{{Vagshette}, N.D.}, \bibinfo{author}{{Pawar}, P.K.}, \bibinfo{author}{{Patil}, M.K.}, \bibinfo{year}{2015}.
\newblock \bibinfo{title}{{X-ray cavities and temperature jumps in the environment of the strong cool core cluster Abell 2390}}.
\newblock \bibinfo{journal}{\apss} \bibinfo{volume}{359}, \bibinfo{pages}{61}.
\newblock \DOIprefix\doi{10.1007/s10509-015-2508-z}, \href{http://arxiv.org/abs/1412.8632}{{\tt arXiv:1412.8632}}.
%Type = Article
\bibitem[{{Spergel} et~al.(2007){Spergel}, {Bean}, {Dor{\'e}}, {Nolta}, {Bennett}, {Dunkley}, {Hinshaw}, {Jarosik}, {Komatsu}, {Page}, {Peiris}, {Verde}, {Halpern}, {Hill}, {Kogut}, {Limon}, {Meyer}, {Odegard}, {Tucker}, {Weiland}, {Wollack} and {Wright}}]{2007ApJS..170..377S}
\bibinfo{author}{{Spergel}, D.N.}, \bibinfo{author}{{Bean}, R.}, \bibinfo{author}{{Dor{\'e}}, O.}, \bibinfo{author}{{Nolta}, M.R.}, \bibinfo{author}{{Bennett}, C.L.}, \bibinfo{author}{{Dunkley}, J.}, \bibinfo{author}{{Hinshaw}, G.}, \bibinfo{author}{{Jarosik}, N.}, \bibinfo{author}{{Komatsu}, E.}, \bibinfo{author}{{Page}, L.}, \bibinfo{author}{{Peiris}, H.V.}, \bibinfo{author}{{Verde}, L.}, \bibinfo{author}{{Halpern}, M.}, \bibinfo{author}{{Hill}, R.S.}, \bibinfo{author}{{Kogut}, A.}, \bibinfo{author}{{Limon}, M.}, \bibinfo{author}{{Meyer}, S.S.}, \bibinfo{author}{{Odegard}, N.}, \bibinfo{author}{{Tucker}, G.S.}, \bibinfo{author}{{Weiland}, J.L.}, \bibinfo{author}{{Wollack}, E.}, \bibinfo{author}{{Wright}, E.L.}, \bibinfo{year}{2007}.
\newblock \bibinfo{title}{{Three-Year Wilkinson Microwave Anisotropy Probe (WMAP) Observations: Implications for Cosmology}}.
\newblock \bibinfo{journal}{\apjs} \bibinfo{volume}{170}, \bibinfo{pages}{377--408}.
\newblock \DOIprefix\doi{10.1086/513700}, \href{http://arxiv.org/abs/astro-ph/0603449}{{\tt arXiv:astro-ph/0603449}}.
%Type = Article
\bibitem[{{Tittley} and {Henriksen}(2005)}]{2005ApJ...618..227T}
\bibinfo{author}{{Tittley}, E.R.}, \bibinfo{author}{{Henriksen}, M.}, \bibinfo{year}{2005}.
\newblock \bibinfo{title}{{Cluster Mergers, Core Oscillations, and Cold Fronts}}.
\newblock \bibinfo{journal}{\apj} \bibinfo{volume}{618}, \bibinfo{pages}{227--236}.
\newblock \DOIprefix\doi{10.1086/425952}, \href{http://arxiv.org/abs/astro-ph/0409177}{{\tt arXiv:astro-ph/0409177}}.
%Type = Article
\bibitem[{{Vagshette} et~al.(2017){Vagshette}, {Naik}, {Patil} and {Sonkamble}}]{2017MNRAS.466.2054V}
\bibinfo{author}{{Vagshette}, N.D.}, \bibinfo{author}{{Naik}, S.}, \bibinfo{author}{{Patil}, M.K.}, \bibinfo{author}{{Sonkamble}, S.S.}, \bibinfo{year}{2017}.
\newblock \bibinfo{title}{{Detection of a pair of prominent X-ray cavities in Abell 3847}}.
\newblock \bibinfo{journal}{\mnras} \bibinfo{volume}{466}, \bibinfo{pages}{2054--2066}.
\newblock \DOIprefix\doi{10.1093/mnras/stw3227}, \href{http://arxiv.org/abs/1612.02560}{{\tt arXiv:1612.02560}}.
%Type = Article
\bibitem[{{Vagshette} et~al.(2016){Vagshette}, {Sonkamble}, {Naik} and {Patil}}]{2016MNRAS.461.1885V}
\bibinfo{author}{{Vagshette}, N.D.}, \bibinfo{author}{{Sonkamble}, S.S.}, \bibinfo{author}{{Naik}, S.}, \bibinfo{author}{{Patil}, M.K.}, \bibinfo{year}{2016}.
\newblock \bibinfo{title}{{AGN-driven perturbations in the intracluster medium of the cool-core cluster ZwCl 2701}}.
\newblock \bibinfo{journal}{\mnras} \bibinfo{volume}{461}, \bibinfo{pages}{1885--1897}.
\newblock \DOIprefix\doi{10.1093/mnras/stw1420}, \href{http://arxiv.org/abs/1606.04651}{{\tt arXiv:1606.04651}}.
%Type = Article
\bibitem[{{van den Bosch} et~al.(2005){van den Bosch}, {Weinmann}, {Yang}, {Mo}, {Li} and {Jing}}]{2005MNRAS.361.1203V}
\bibinfo{author}{{van den Bosch}, F.C.}, \bibinfo{author}{{Weinmann}, S.M.}, \bibinfo{author}{{Yang}, X.}, \bibinfo{author}{{Mo}, H.J.}, \bibinfo{author}{{Li}, C.}, \bibinfo{author}{{Jing}, Y.P.}, \bibinfo{year}{2005}.
\newblock \bibinfo{title}{{The phase-space parameters of the brightest halo galaxies}}.
\newblock \bibinfo{journal}{\mnras} \bibinfo{volume}{361}, \bibinfo{pages}{1203--1215}.
\newblock \DOIprefix\doi{10.1111/j.1365-2966.2005.09260.x}, \href{http://arxiv.org/abs/astro-ph/0502466}{{\tt arXiv:astro-ph/0502466}}.
%Type = Article
\bibitem[{{Vikhlinin} et~al.(2001){Vikhlinin}, {Markevitch} and {Murray}}]{2001ApJ...551..160V}
\bibinfo{author}{{Vikhlinin}, A.}, \bibinfo{author}{{Markevitch}, M.}, \bibinfo{author}{{Murray}, S.S.}, \bibinfo{year}{2001}.
\newblock \bibinfo{title}{{A Moving Cold Front in the Intergalactic Medium of A3667}}.
\newblock \bibinfo{journal}{\apj} \bibinfo{volume}{551}, \bibinfo{pages}{160--171}.
\newblock \DOIprefix\doi{10.1086/320078}, \href{http://arxiv.org/abs/astro-ph/0008496}{{\tt arXiv:astro-ph/0008496}}.
%Type = Article
\bibitem[{{Walker} et~al.(2014){Walker}, {Fabian} and {Sanders}}]{2014MNRAS.441L..31W}
\bibinfo{author}{{Walker}, S.A.}, \bibinfo{author}{{Fabian}, A.C.}, \bibinfo{author}{{Sanders}, J.S.}, \bibinfo{year}{2014}.
\newblock \bibinfo{title}{{Large-scale gas sloshing out to half the virial radius in the strongest cool core REXCESS galaxy cluster, RXJ2014.8-2430.}}
\newblock \bibinfo{journal}{\mnras} \bibinfo{volume}{441}, \bibinfo{pages}{L31--L35}.
\newblock \DOIprefix\doi{10.1093/mnrasl/slu040}, \href{http://arxiv.org/abs/1402.6894}{{\tt arXiv:1402.6894}}.
%Type = Article
\bibitem[{{Wilms} et~al.(2000){Wilms}, {Allen} and {McCray}}]{2000ApJ...542..914W}
\bibinfo{author}{{Wilms}, J.}, \bibinfo{author}{{Allen}, A.}, \bibinfo{author}{{McCray}, R.}, \bibinfo{year}{2000}.
\newblock \bibinfo{title}{{On the Absorption of X-Rays in the Interstellar Medium}}.
\newblock \bibinfo{journal}{\apj} \bibinfo{volume}{542}, \bibinfo{pages}{914--924}.
\newblock \DOIprefix\doi{10.1086/317016}, \href{http://arxiv.org/abs/astro-ph/0008425}{{\tt arXiv:astro-ph/0008425}}.
%Type = Article
\bibitem[{{Wu} et~al.(2000){Wu}, {Fabian} and {Nulsen}}]{2000MNRAS.318..889W}
\bibinfo{author}{{Wu}, K.K.S.}, \bibinfo{author}{{Fabian}, A.C.}, \bibinfo{author}{{Nulsen}, P.E.J.}, \bibinfo{year}{2000}.
\newblock \bibinfo{title}{{Non-gravitational heating in the hierarchical formation of X-ray clusters}}.
\newblock \bibinfo{journal}{\mnras} \bibinfo{volume}{318}, \bibinfo{pages}{889--912}.
\newblock \DOIprefix\doi{10.1046/j.1365-8711.2000.03828.x}, \href{http://arxiv.org/abs/astro-ph/9907112}{{\tt arXiv:astro-ph/9907112}}.

\end{thebibliography}



%% For example:

%% \bibitem[Aladro et al.(2015)]{Aladro15} Aladro, R., Martín, S., Riquelme, D., et al. 2015, \aas, 579, A101

%%\end{thebibliography}

\end{document}